\documentclass[twocolumn]{aastex7}

\usepackage{graphicx}
\usepackage{longtable}
\usepackage{mathtools}
\usepackage{bm}
\usepackage{booktabs}
\usepackage{multirow}
\usepackage{float}
\usepackage{hyperref}
\usepackage{url}
\usepackage{wrapfig}
\definecolor{crimson}{RGB}{230, 20, 40}

\begin{document}

\title{JWST/NIRSpec Reveals Diverse Nuclear Environments in Dwarf Galaxies Hosting AGN}

\accepted{September 3, 2026}

\submitjournal{ApJ}

\author[0000-0002-4375-254X]{Thomas Bohn}
\affil{Institute for Academic Innovation, Hokkaido University, N10 W8 Kitaku, Sapporo, Hokkaido, 060-0810, Japan}
\email[show]{tbohn002@ucr.edu}

\author[0000-0001-7578-2412]{Archana Aravindan}
\affiliation{Department of Astronomy, The University of Texas at Austin, Austin, TX 78712, USA}
\affiliation{Cosmic Frontier Center, The University of Texas at Austin, Austin, TX 78712, USA}
\email{}

\author[0000-0003-4693-6157]{Gabriela Canalizo}
\affiliation{Department of Physics and Astronomy, University of California, Riverside, 900 University Ave, Riverside CA 92521, USA}
\email{}

\author[0009-0003-7749-1864]{Aditya Togi}
\affiliation{Department of Physics, Texas State University, 601 University Drive, San Marcos, TX 78666, USA}
\email{}

\author[0000-0003-2277-2354]{Shobita Satyapal}
\affiliation{Department of Physics and Astronomy, George Mason University, MS3F3, 4400 University Drive, Fairfax, VA 22030, USA}
\email{}

\author[0000-0002-7402-5441]{Tohru Nagao}
\affiliation{Ehime University, Bunkyo-cho 2-5, Matsuyama, Ehime 790-8577, Japan}
\email{}

\author[0000-0001-6919-1237]{Matthew Malkan}
\affiliation{Division of Astronomy and Astrophysics, University of California, Los Angeles, CA 90095}
\email{}

\author[0000-0001-8490-6632]{Thomas S.Y. Lai}
\affiliation{IPAC, MC 320-6, Caltech, 1200 E. California Blvd., Pasadena, CA 91125}
\email{}

\author[0000-0003-4268-0393]{Hanae Inami}
\affiliation{Hiroshima University, 1-3-1 Kagamiyama, Higashi-Hiroshima City, Hiroshima, 739-8526, Japan}
\email{}

\author[0000-0003-3762-7344]{Weizhe Liu}
\affiliation{Steward Observatory, University of Arizona, 933 N. Cherry Ave,
Tucson, AZ 85721, USA}
\email{}

\author[0000-0002-6570-9446]{Marina Bianchin}
\affiliation{Instituto de Astrof\'{\i}sica de Canarias, Calle V\'{\i}a L\'{a}ctea s/n, E-38205, La Laguna, Tenerife, Spain}
\affiliation{Departamento de Astrof\' isica, Universidad de La Laguna, E-38206, La Laguna, Tenerife, Spain}
\email{}

\author[0000-0003-0699-6083]{Tanio Diaz-Santos}
\affiliation{Institute of Astrophysics - FORTH, GR-70013 Vassilika Vouton,
Greece}
\affiliation{School of Sciences, European University Cyprus, Diogenes street, Engomi, 1516 Nicosia, Cyprus}
\email{}

\author[0000-0003-2638-1334]{Aaron Evans}
\affiliation{Department of Astronomy, University of Virginia, 530 McCormick Road, Charlottesville, VA 22904
NRAO, 520 Edgemont Road, Charlottesville, VA, 22903}
\email{}

\author[0000-0002-1000-6081]{Sean T. Linden}
\affiliation{Steward Observatory, University of Arizona, 933 N. Cherry Ave,
Tucson, AZ 85721, USA}
\email{}

\author[0000-0002-1912-0024]{Vivian U}
\affiliation{IPAC, California Institute of Technology, 1200 E. California Blvd., Pasadena, CA 91125, USA}
\email{}

\author[0000-0003-3498-2973]{Lee Armus}
\affiliation{IPAC, MC 320-6, Caltech, 1200 E. California Blvd., Pasadena, CA 91125}
\email{}

\author[0000-0003-0057-8892]{Loreto Barcos-Muñoz}
\affiliation{North American ALMA Science Center, National Radio Astronomy Observatory, 520 Edgemont Road, Charlottesville, VA 22903}
\email{}

\author[0000-0003-3917-6460]{Kirsten Larson}
\affiliation{AURA for the European Space Agency (ESA), Space Telescope Science Institute, 3700 San Martin Drive, Baltimore, MD 21218, USA}
\email{}

\author[0000-0002-3139-3041]{Yiqing Song}
\affiliation{European Southern Observatory, Joint ALMA Observatory, Alonso de Co\'{r}dova, 3107, Vitacura, Santiago, 763-0355, Chile}
\email{}

\author[0000-0002-7532-3328]{Sabrina Stierwalt}
\affiliation{Occidental College, 1600 Campus Road, Los Angeles, CA, 90041}
\email{}

\author[0000-0001-7291-0087]{Jason Surace}
\affiliation{IPAC, MC 314-6, Caltech, 1200 E. California Blvd., Pasadena, CA 91125}
\email{}

\begin{abstract}

Dwarf galaxies, in the $\Lambda$CDM framework, are expected to dominate the galaxy number density at all redshifts. As such, studying AGN in these systems, including 
characterizing their local environments and emission properties, is essential in obtaining a comprehensive view of the AGN population and their influence on their host galaxy. To this end, we present \textit{JWST}/NIRSpec IFU observations of four dwarf galaxies (M$_\star<$10$^{9.5}\;$M$_\odot$) with evidence of AGN activity. Utilizing the improvements in resolution and sensitivity offered by \textit{JWST}, we investigate the emission features of the central kiloparsec of these dwarfs. Our findings include the detection of over 100 emission lines across our sample, including coronal lines with ionization potentials greater than 300 eV. The nuclear spectra show diverse emission features. In two galaxies, we measure strong contributions from hot dust to the infrared continuum. We also detect both PAH and coronal line emission within 100 parsecs of the nuclei in three galaxies. We estimate the hot H$_2$ gas mass to be between 1--20 M$_\odot$ within 300 pc across our sample, where both thermal and non-thermal excitation are involved. Lastly, spatial flux maps reveal varied emission structures across all observed gas phases, ranging from radial symmetry around the nucleus to elongated plumes. Kinematic maps also show each galaxy has unique velocity structures despite all being isolated, affirming the diversity of rotation curves problem, a long standing tension between simulations and observations. These results showcase the unique and varied nuclear environments that can be found in dwarf galaxies hosting AGN.

\end{abstract}






\section{Introduction}

Studies of active galactic nuclei (AGN) in massive galaxies ($M_{\star} >$ 10$^{10}$ $M_{\odot}$) have served as a basis for our understanding of the coevolution between supermassive black holes (SMBHs) and their host galaxies. Well-established relations have emerged as a result, including BH mass with stellar velocity dispersion \citep[e.g.,][]{McConnell2013,Sexton2019} and stellar mass \citep[e.g.,][]{Reines2015}. These relations are often attributed to a combination of gas rich mergers that can fuel AGN growth \citep[e.g.,][]{Hopkins2006,Ricci2017b} and subsequent feedback events that can regulate star formation through kiloparsec (kpc) scale outflows \citep{King2015,Veilleux2020}.

One of the recent developments in this coevolution is the discovery of a new population of lower mass ($\sim$10$^{6}$ - 10$^{8}$ $M_{\odot}$) SMBHs located at $z>4$ \citep[e.g.,][]{Larson2023,Kocevski2023,Harikane2023,Greene2024,Furtak2024,Kokorev2024}. Interestingly, this new population of BHs can be up to 100 times more massive than expected from local stellar mass relations, suggesting a redshift evolution in the $M_{\rm{BH}}$ -- $M_{*}$ relation \citep{Pacucci2024}. Although uncertainties of the involved measurements can be large, these high $M_{\rm{BH}}$/$M_{*}$ ratios could be indicative of massive BH seeds in the early Universe \citep{Greene2020,Natarajan2024}.

While these findings are enhancing our view of the early BH population, we still lack a clear consensus on the preferred BH seeding scenarios and growth mechanisms. A prevalent issue is that direct observations and mass measurements of SMBHs are limited to massive nearby galaxies. Additionally, high redshift studies are often biased towards the most luminous quasars and Type-1 AGN, which represent only a fraction of the total AGN population \citep{Ho2008,Assef2013}.

An alternative avenue to constrain the earliest BH formation scenarios and number densities is through the detection of AGN in dwarf galaxies. Because of their relative quiescent merger histories, low mass galaxies are more suited to study secular BH growth mechanisms, allowing us to more directly constrain seeding models \citep{Greene2020}. Indeed, the number of AGN candidates in dwarfs has seen substantial growth in the past decade \citep{Reines2013,Moran2014,Sartori2015,Burke2024,Wasleske2024}. Additionally, dwarfs below 10$^{9}$ $M_{\odot}$ are expected to host BHs in the intermediate mass range (IMBH, $\sim$10$^{2}$ -- 10$^{5}$ $M_{\odot}$), an important population whose detections will help establish a holistic view of BH-galaxy coevolution. While some candidate IMBHs have been detected \citep{Cann2018,Chisholm2024}, there are currently no secure mass measurements of BHs in this mass regime.

While the IMBH population remains elusive, progress has been made in confirming the presence of AGN in dwarfs and evaluating their impact on their host galaxy \citep{Manzano2019,Liu2020,Aravindan2023,Liu2024,Salehirad2025}. One of the most surprising outcomes of these studies is the presence of ionized outflowing gas that is likely originating from the AGN. While intriguing, the number of confirmed AGN outflows in dwarfs is small and, on a more fundamental level, the properties and gas dynamics of dwarf galaxies hosting AGN remain poorly understood. As a result, BH physics as a whole are not widely included in dwarf galaxy simulations \citep{Sales2022}, an inclusion that could have important implications on long standing discrepancies between observations and simulations, including the core-cusp problem \citep{Flores1994,Adams2014} and rotation curve diversity \citep{Oman2015}. Put together, the prospects of finding IMBHs, characterizing the conditions of the nuclear environments, and reexamining the role of AGN feedback in dwarfs necessitate a closer look at these systems. 

In this article, we present \textit{James Webb Space Telescope} (\textit{JWST}) NIRSpec Integral Field Spectroscopy (IFS) observations of four dwarf galaxies with evidence of AGN activity. Despite being isolated systems with no indication of merger activity, we find substantial diversity in the continuum shape, coronal line (CL) detections, polycyclic aromatic hydrocarbon (PAH) emission, molecular H$_2$ gas excitation, and gas morphology. The purpose of this paper is to provide an overview of the data, detailing the unique and varied characteristics of these compelling systems. It is organized as follows: in Section \ref{sec:Sample_Obs}, we provide details of the sample selection, observations, and data reduction. Section \ref{sec:Analysis} describes the emission line fitting process and spatial map creation, the results of which are shared in Section \ref{sec:Results}. Section \ref{sec:Discussion} concludes with discussing the unique ionized and molecular emission features found within our sample. For details regarding the highly ionized outflows, we refer to the reader to an accompanying paper \citep{Aravindan2026}.

Throughout our analysis, we adopt a cosmology of $H_0=70$ km s$^{-1}$ Mpc$^{-1}$, $\Omega_\Lambda=0.70$, and $\Omega_{\rm matter}=0.30$.

\begin{deluxetable}{cccccccc}
\setlength{\tabcolsep}{8pt}
\caption{Summary of Observations} 
\label{tab:Sample_obs}
\tablehead{\colhead{Galaxy} & \colhead{RA} & \colhead{Dec} & \colhead{Redshift} & \colhead{Log($M_{*}/M_\odot$)}  & \colhead{Pixel Scale} & \colhead{Exposure Time}\\
\colhead{} & \colhead{(J2000)} & \colhead{(J2000)} & \colhead{} & \colhead{} & \colhead{(pc/$0\farcs05$)} & \colhead{(s)}}
\startdata
J1009 & 10:09:35.66 & 26:56:48.93 & 0.0145 & 8.77  & 15 & 3,500\\
J0954 & 09:54:18.16 & 47:17:25.08 & 0.0329 & 9.12 & 33 & 2,900\\
J0842 & 08:42:34.51 & 03:19:30.69 & 0.0291 & 9.34  & 30 & 3,500\\
J0906 & 09:06:13.75 & 56:10:15.13 & 0.0467 & 9.36  & 46 & 2,900
\enddata
\tablecomments{Columns: (1) Short name referenced in this study, in order of increasing galaxy stellar mass. (2-3) Right Ascension and Declination of each galaxy. (4) Redshift, as calculated from fits to stellar absorption features. (5) Galaxy stellar mass from the NASA-Sloan Atlas (NSA\footnote{\url{https://nsatlas.org/}}). (6) Pixel size ($0\farcs05$) in parsecs. (7) Total on-source exposure time for each filter in seconds.}
\end{deluxetable}

\section{Sample and Observations} \label{sec:Sample_Obs}

\subsection{Sample Selection} \label{subsec:Sample}

The four dwarf galaxies that serve as the focus of this article were first identified as AGN candidates in a series of surveys aimed at identifying AGN in low mass hosts \citep{Reines2013,Moran2014,Sartori2015}. Here, hundreds of AGN candidates were identified based on diagnostics including MIR color cuts and BPT \citep{Baldwin1981} optical line ratios (\ion{He}{2}, [\ion{O}{1}], and [\ion{S}{2}]). Drawing from these parent samples, \cite{Manzano2019} obtained Keck/LRIS observations of 29 of these candidates, where $\sim$30$\%$ exhibited broad, asymmetric [\ion{O}{3}] $\lambda$5007 emission profiles, with average $W_{80, \rm{broad}}$\footnote[1]{Line width containing 80$\%$ of the emission line flux.} exceeding 1,000 km s$^{-1}$, a tell-tale sign of outflows. Follow-up IFU observations conducted by \cite{Liu2020} with Keck/KCWI confirmed ionized outflows in 7/8 targets (with velocities up to 1,200 km s$^{-1}$), with five extending out to a kpc or more. Furthermore, \cite{Bohn2021} detected near-infrared (NIR) CL\footnote[2]{Coronal lines (CLs) are highly ionized, forbidden transitions with ionization potentials greater than what is typically produced by stellar processes. Their presence strongly indicates AGN activity.} emission in 5/9 galaxies, and found a higher incidence of CL detections with targets hosting more energetic outflows. 

\begin{figure}
\centering
\epsscale{1.2}
\plotone{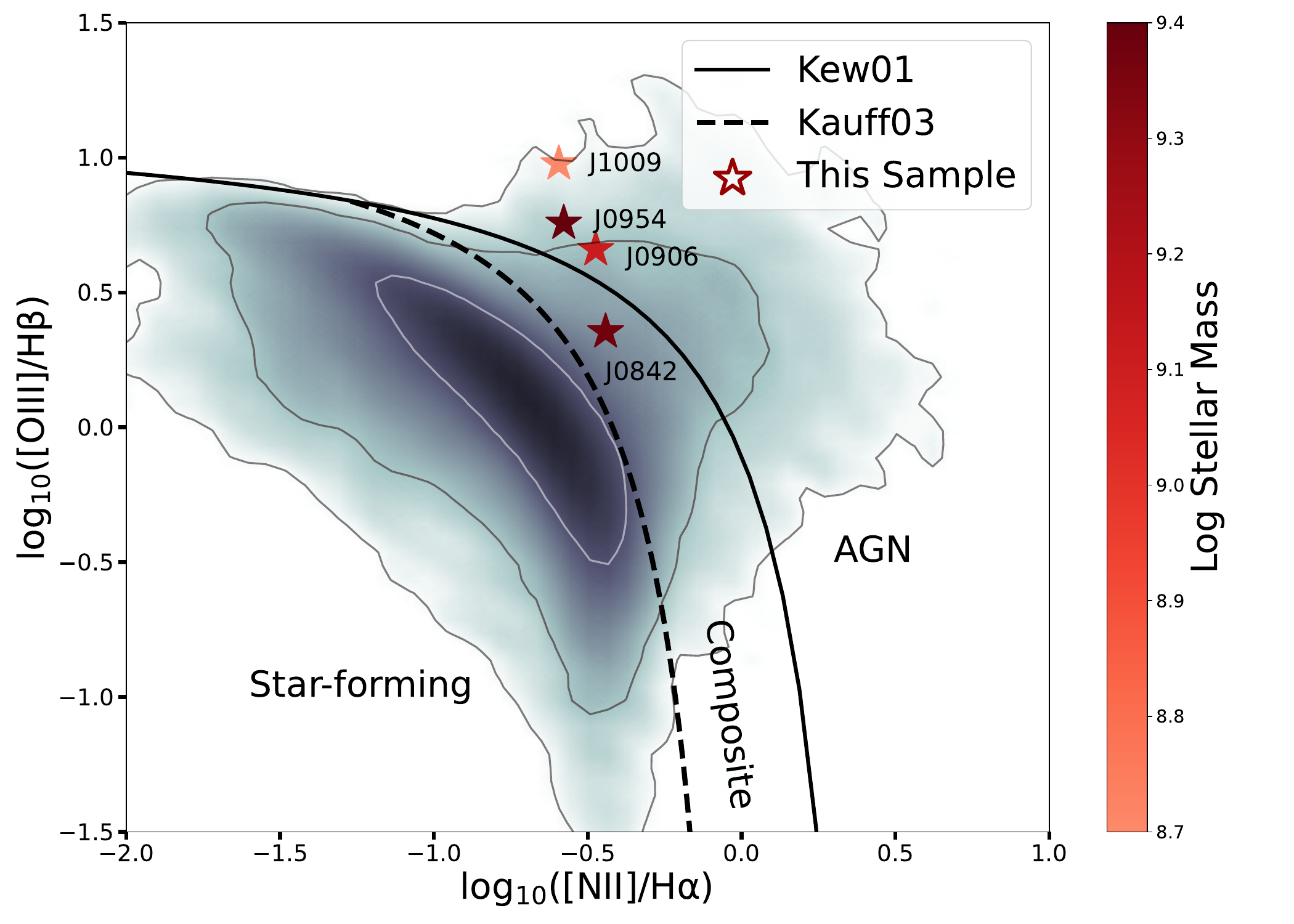}
\caption{BPT line ratios of the central 200 pc emission based on Keck/LRIS spectroscopy (see \cite{Manzano2019}). Galaxies with stellar masses $M_{*}<10^{9.5}\; \rm{M_{\odot}}$ from the full NASA-Sloan Atlas (NSA) are represented as the shaded region, with contours representing 1$\sigma$, 2$\sigma$ and 3$\sigma$ levels. Only one target, J0842, is located in the composite region while the rest show line ratios consistent with AGN.
\label{fig:BPT}}
\end{figure}

With these previous results, we constructed our sample from \cite{Manzano2019}, where we first imposed a mass cutoff of $<$10$^{9.5}$ $M_{\odot}$. Of the remaining eight targets, we selected those with NIR CL emission, as reported in \cite{Bohn2021}, resulting in four dwarf galaxies (see Table \ref{tab:Sample_obs}). These four cover a narrow range of stellar masses (0.6 dex), with the lowest being 10$^{8.8}$ $M_{\odot}$. They also span a relatively small region of the BPT diagram, where line ratios differ by only 0.8 dex (Figure \ref{fig:BPT}). Based on previous Keck observations, all four have broad [\ion{O}{3}] $\lambda$5007 subcomponents with estimated velocity offsets and $W_{80}$ line widths as high as 220 and 1250 km s$^{-1}$, respectively \citep{Liu2020}. As mentioned previously, these systems are all isolated with only two having $\sim$5 neighbors of comparable mass within 1.5 Mpc. Given these distinctive properties among dwarf galaxies, these four were selected for follow-up NIRSpec IFU observations with \textit{JWST} to investigate their central ionized and molecular emission properties.

\subsection{Observations and Data Reduction} \label{subsec:Observation}

The data were obtained between March and May 2024, as part of the \textit{JWST} program PID 3663 (P.I. Bohn). Observations were taken with NIRSpec IFS \citep{Jakobsen2022,Boker2022}, where three high resolution gratings were utilized, G140H/F100LP, G235H/F170LP, and G395H/F290LP. This setup resulted in a wavelength coverage of 0.97--5.27 $\mu$m and a mean nominal resolving power of R$\sim$2700 (110 km s$^{-1}$), where the accuracy of the NIRSpec wavelength calibration is 1/8 of the resolution element or better ($<$14 km s$^{-1}$, see \cite{Boker2023} for further details). A four-point dither pattern was used for all gratings to improve the PSF sampling. Background images were also obtained for each galaxy to reduce the background level in order to detect faint emission lines. Details of the observations are summarized in Table \ref{tab:Sample_obs}.

We downloaded the uncalibrated science and background frames via the MAST portal and reprocessed the data through version 1.15.1 of the \textit{JWST} Science Calibration Pipeline \citep{Bushouse2023}, with CRDS context jwst1253.pmap. We followed the standard three step reduction process of the pipeline. Briefly, the first stage, \textsc{DETECTOR1}, applies detector-level corrections to the uncalibrated frames, from which 2D count-rate images are made. Notably, we applied the `snowball' rejection algorithm to flag instances of energetic cosmic ray events. Stage 2 (\textsc{spec2}) implements various calibration steps, including distortion, wavelength, and flux calibrations. Background subtraction was also performed using dedicated background frames during this step. The final datacubes were generated during stage 3 (\textsc{spec3}), at which we opted to change the spatial scaling to $0\farcs05$ to enhance the spatial sampling. 

Inspection of the final pipeline products revealed sinusoidal variations in the continuum, caused by improper tracing of the target during rectification. To correct these modulations, we applied a custom algorithm that follows the method detailed in \cite{Perna2023}. Here, after masking the detected emission lines, we modeled the sinusoidal variations and then subtracted them from the continuum (for further details, see W. Matzko et al. 2026, in prep). Following this correction, any remaining pixel outliers that remained in the post-pipeline reduction were then removed using the sigma-clipping code presented in \cite{Hutchison2024}.

\section{Methods of Analysis} \label{sec:Analysis}

\begin{figure*}
\epsscale{1.15}
\centering
\plotone{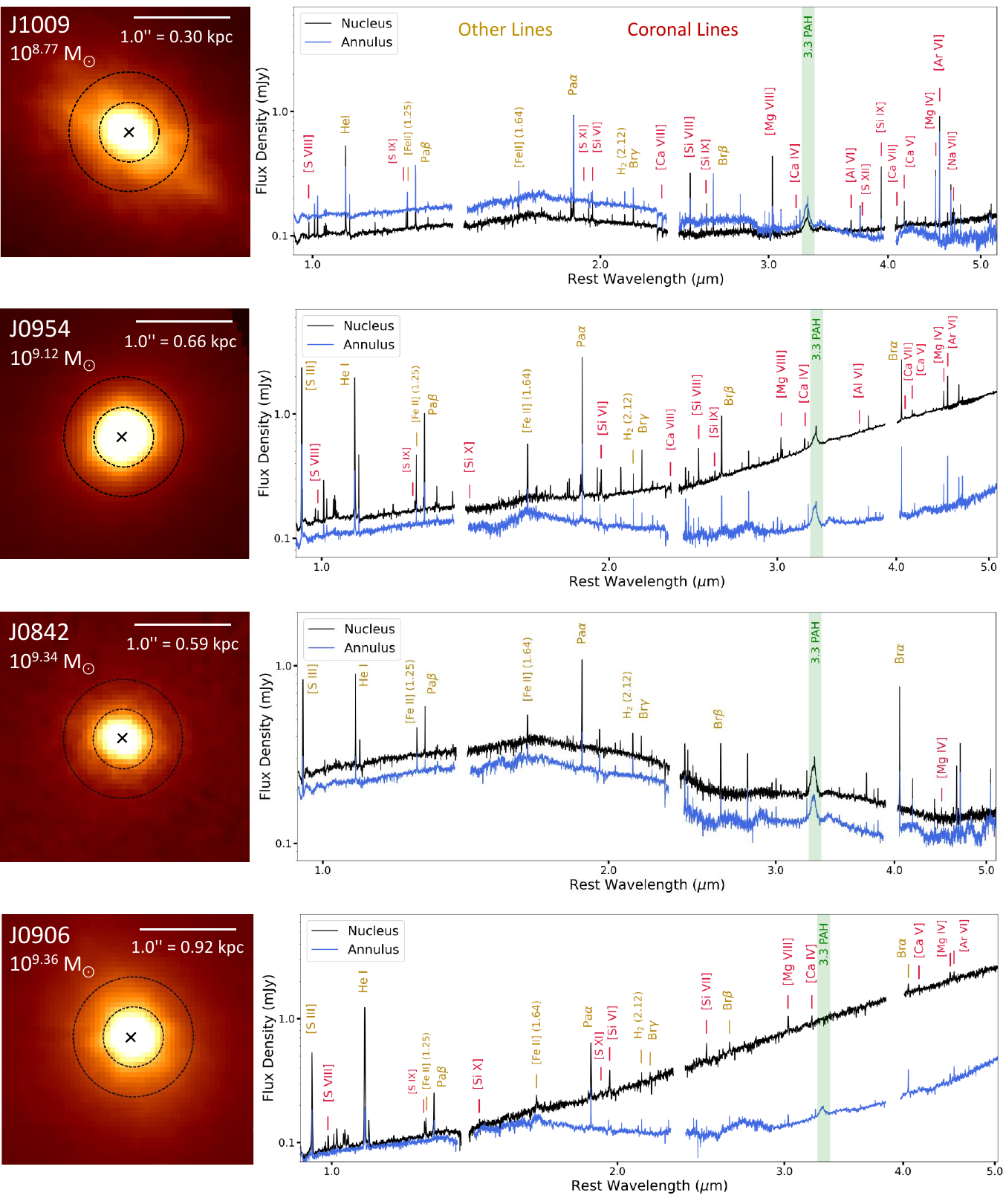}
\caption{(\textit{left}) NIRSpec 2.0--2.1$\mu$m continuum images of each galaxy in this study, in order of increasing stellar mass. Two spectral extractions were taken: a circular aperture of r=$0\farcs3$ centered on the nucleus (marked as a black cross), and a ring (annulus) extraction with r$\rm{_{outer}}$=$0\farcs6$. (\textit{right}) Prominent emission lines, all detected coronal lines, and the 3.3$\mu$m PAH emission are labeled. Among these four dwarfs, we see a diverse set of continuum profiles, both between each galaxy and the inner and outer extractions. \label{fig:spectra}}
\end{figure*}

Two aperture extractions were performed for each target: a $0\farcs3$ radius circular aperture centered on the nucleus, defined as the location of the peak continuum emission, and an encompassing annular extraction with r$\rm{_{inner}}$=$0\farcs3$ and r$\rm{_{outer}}$=$0\farcs6$, as shown in Figure \ref{fig:spectra}. These aperture sizes were selected to limit aperture corrections while also isolating the nuclear emission. 

We fit the spectra using two independent algorithms, each optimized for different fitting purposes. First, we used the Continuum And Feature Extraction (\textsc{CAFE}) tool \citep{Marshall2007,DiazSantos2025} to model the nuclear continuum and assess the relative contributions from the dust, stellar, and AGN components. To fit the NIR continuum, \textsc{CAFE} uses contributions from hot ($T_{hot} \approx$ 1400 K) and warm ($T_{warm} \approx$ 200 K) dust, the AGN accretion disk (modeled as multiple power laws), and starburst emission from 10 and 100 Myr templates based on \textsc{Starburst99} \citep{Leitherer1999}. Parameters of each of these variables include flux, obscuration, and temperature (for the dust components). Importantly, we lack data redwards of 5 $\mu$m and thus we cannot robustly constrain the various contributions. We fit the continuum spectra multiple times with \textsc{CAFE}, altering the initial parameters of each variable between runs but, as expected, degeneracies arise in the temperatures and fluxes of the dust components. Consequently, our aim with \textsc{CAFE} is to only obtain broad estimates of the relative contributions to the nuclear continuum emission.

The other fitting tool we used is the Bayesian AGN Decomposition Analysis for Sloan Digital Sky Survey (SDSS) Spectra \citep[\textsc{BADASS};][]{Sexton2021}, which we used to fit the multitude of emission lines detected in our spectra. Through Markov Chain Monte Carlo routines, \textsc{BADASS} is optimized to perform simultaneous multicomponent fits to emission-line spectra while also obtaining robust error estimates. Since the main purpose of \textsc{BADASS} is to obtain line fluxes, we fit the local continuum with a third-order Legendre polynomial along with stellar template fits to the host galaxy and stellar absorption features. A single systemic velocity, which serves as the reference point for our emission line kinematics, was derived from the median of the H-band stellar templates over our full FoV (see \cite{Aravindan2026} for further details). Errors from the template fitting range from 20-30 km s$^{-1}$ across our sample. While the 2.3 $\mu$m CO-bands generally provide a more robust measurement of the stellar velocity component, most in our data are significantly affected by the NIRSpec detector gap which compromises their reliability.


Emission lines were fit with \textsc{BADASS} using either Gaussian (recombination, molecular, and fine-structure lines) or Lorentzian functions (3.3$\mu$m PAH). If the residuals of a single Gaussian fit to an emission line were significant, we included a second Gaussian. The inclusion of this secondary component was based on the \textit{F}-test of variances: \textit{F} = $(\sigma\rm{_{single}})^{2}/(\sigma\rm{_{double}})^{2}$, where $\sigma$ is the standard deviation of the residuals using either a single or double Gaussian fit. If \textit{F} $>$ 3.0, then adding a second component provided a sufficient improvement to the fit and was thus justifiable. While the parameters of this second component were left unrestricted, the results of the best fit models all settled on larger velocity dispersion for this component, and thus henceforth will be referred to as the `broad' component. For a robust detection, we set a signal-to-noise ratio (S/N) threshold of three for all lines and components. Additionally, if a weaker line was detected close to this threshold (2.5 $<$ S/N $<$ 3.0), we report the line flux as an upper limit. Uncertainties were derived from the error extension of the NIRSpec datacubes and the random errors associated with the Bayesian fitting process. All detected emission lines, including those with upper limits or broad components, are compiled in Table \ref{tab:Flux_table} (see Appendix).

\begin{figure*}
\epsscale{1.15}
\centering
\plotone{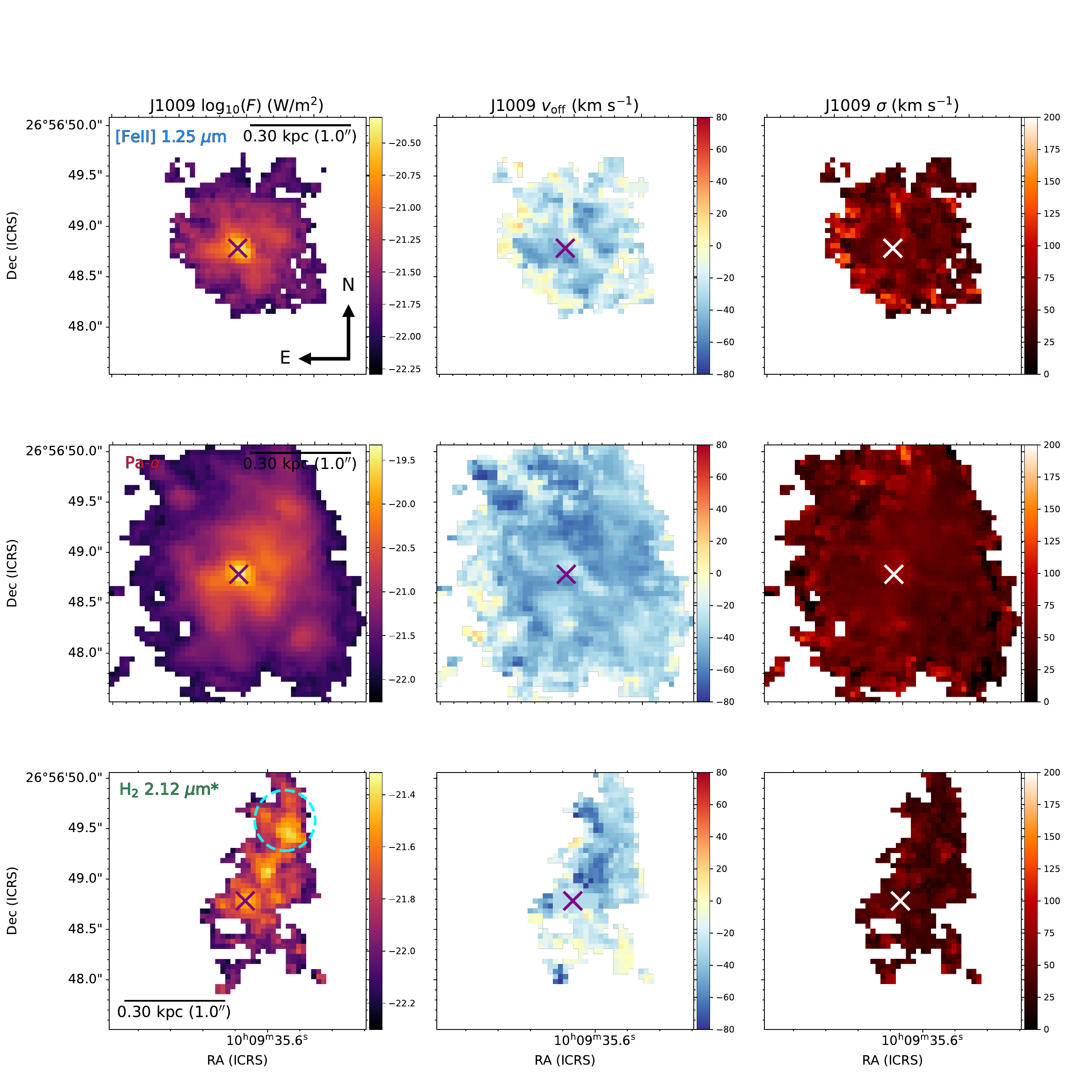}
\caption{J1009 flux, velocity offset, and dispersion maps for [\ion{Fe}{2}] 1.25$\mu$m, Pa-$\alpha$, and H$_2$ 2.12$\mu$m based on single Gaussian fits. Each spaxel was fit iteratively using one Gaussian for each emission line, where a S/N detection threshold of 3 was set. v$\rm{_{off}}$ is defined as the offset from the systemic velocity of the nucleus, as calculated from stellar absorption features. The location of the nucleus is shown as a purple (or white) cross. All panels are oriented such that North is up and East is left. Due to the undersampling of the PSF, individual spaxel fits to H$_2$ 2.12 $\mu$m could not be properly done within the nucleus. A box extraction was used to replace the affected spaxels (see Section \ref{sec:Analysis} for further details), where an asterisk next to the line name denotes affected emission maps. An additional extraction sampling the NW emission plume is shown as a dotted cyan circle (see Section \ref{subsubsec:substructure}). \label{fig:J1009_maps}}
\end{figure*}

\begin{figure*}
\epsscale{1.15}
\centering
\plotone{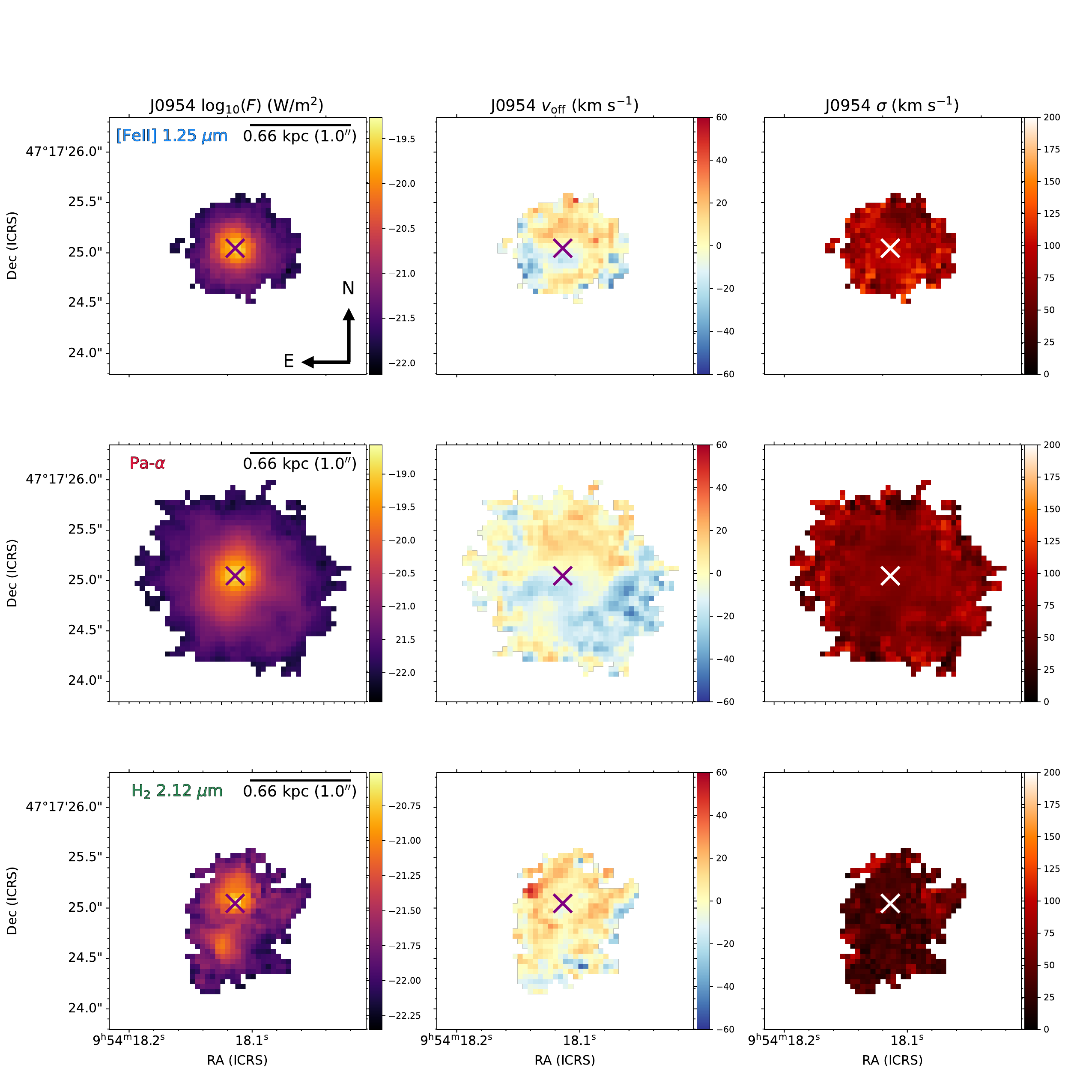}
\caption{Same as Figure \ref{fig:J1009_maps}, but for J0954. \label{fig:J00954_maps}}
\end{figure*}

\begin{figure*}
\epsscale{1.15}
\centering
\plotone{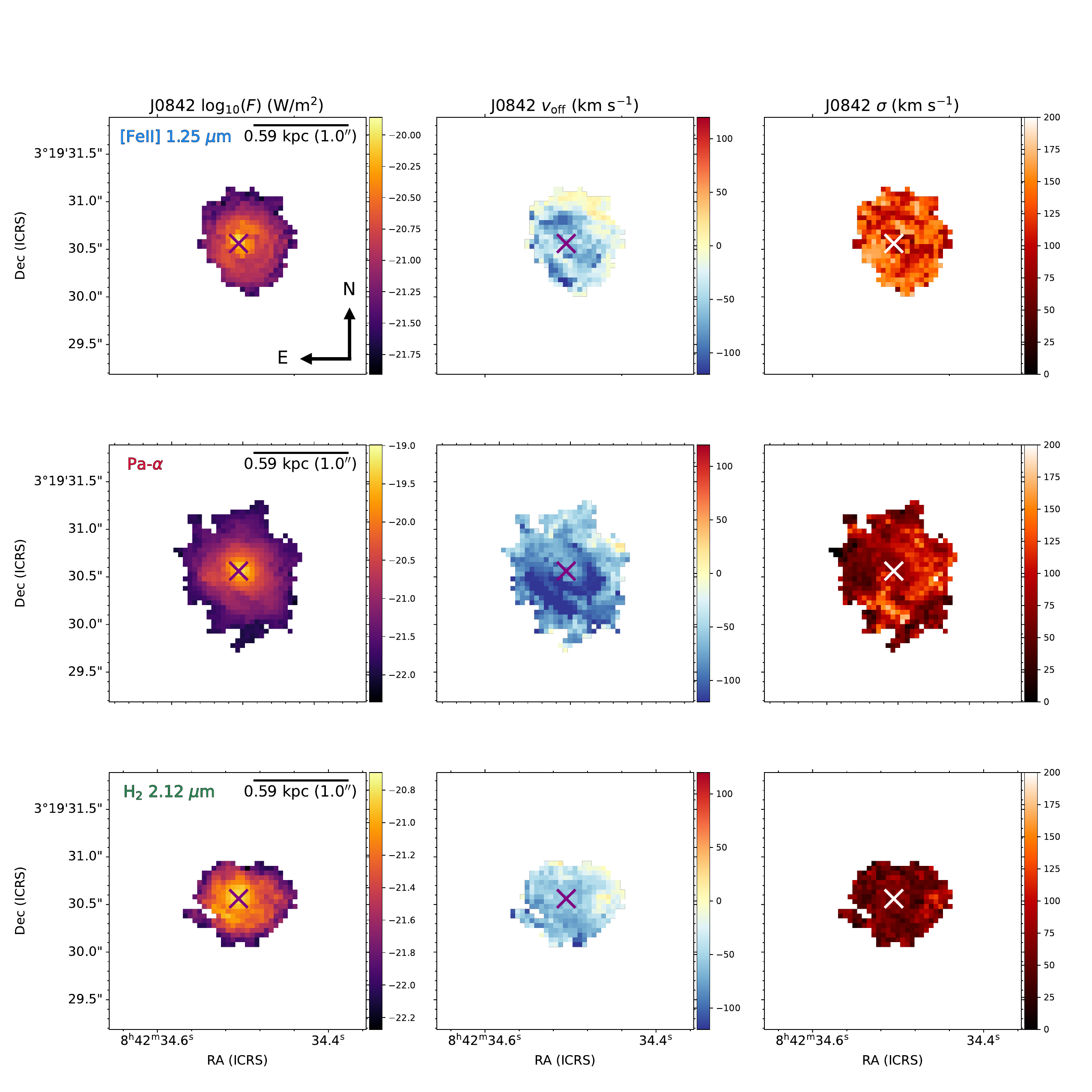}
\caption{Same as Figure \ref{fig:J1009_maps}, but for J0842. \label{fig:J0842_maps}}
\end{figure*}

\begin{figure*}
\epsscale{1.15}
\centering
\plotone{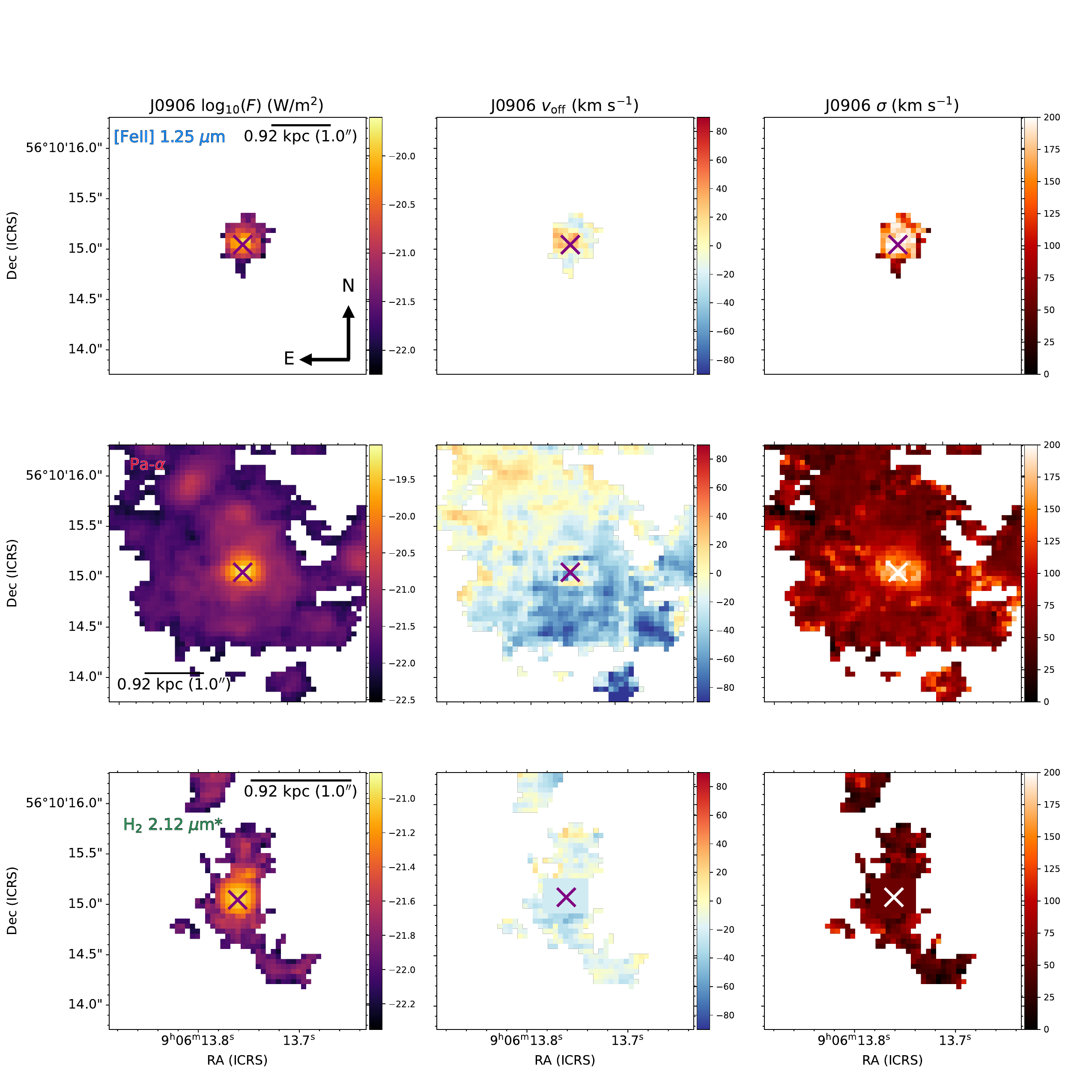}
\caption{Same as Figure \ref{fig:J1009_maps}, but for J0906. Similar to J1009, the H$_2$ 2.12 $\mu$m emission in the nucleus was undersampled. For details regarding the box extraction used to replace the affected spaxels, see Section \ref{sec:Analysis}.\label{fig:J0906_maps}}
\end{figure*}

\begin{figure*}
\epsscale{1.15}
\centering
\plotone{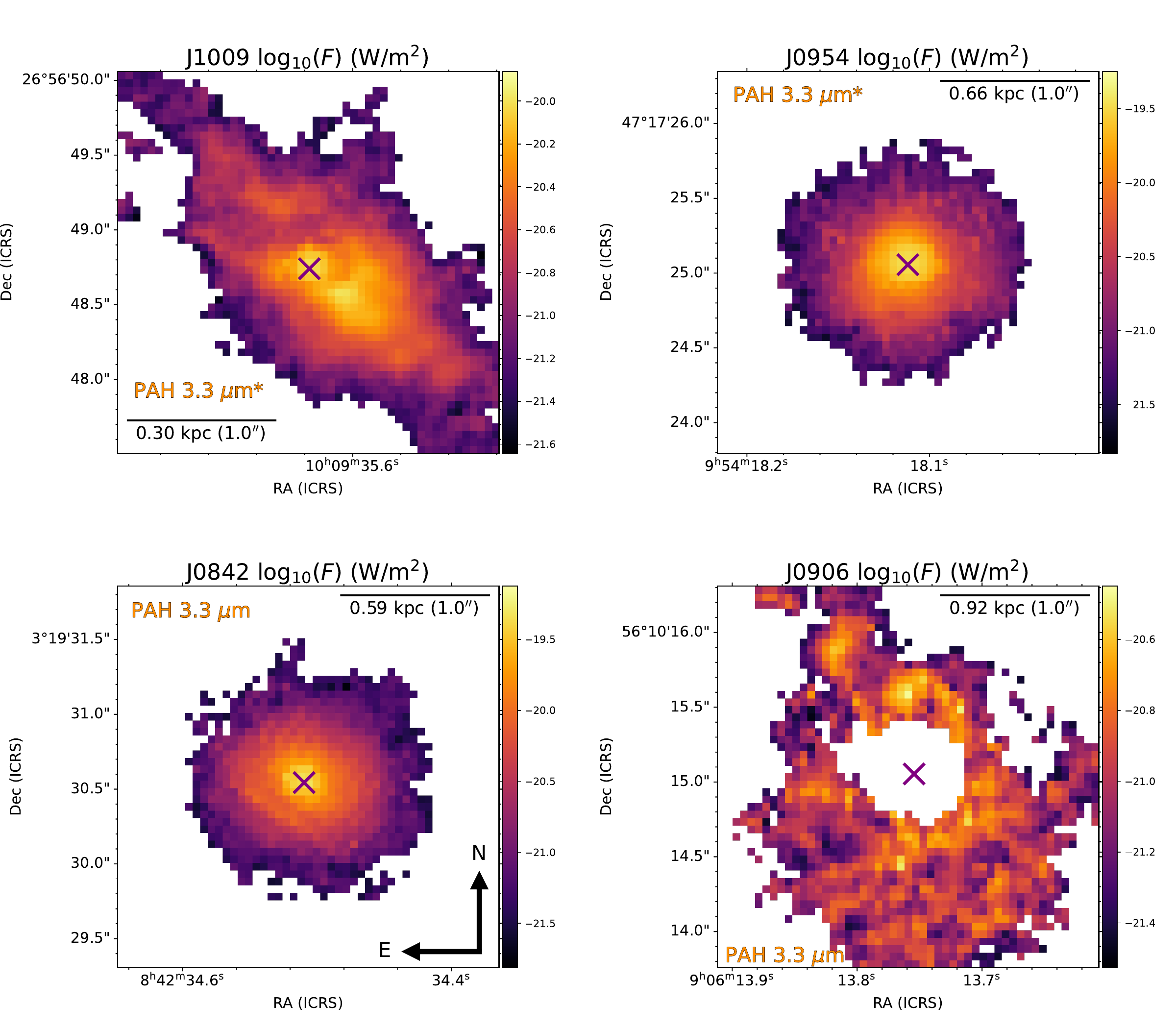}
\caption{PAH 3.3$\mu$m flux maps for each galaxy, where a single Lorentzian was used for each fit. The location of the nucleus is shown as a purple cross, and North/East are up/right in all panels. For J0954 and J1009, robust fits to the PAH emission on the individual spaxel level could not be done due to undersampling the PSF. A box extraction was used to replace the affected spaxels, as described in Section \ref{sec:Analysis}. \label{fig:PAH_maps}}
\end{figure*}

Another key feature of \textsc{BADASS} is the ability to create spatial maps based on individual spaxel fits. The spectra of each spaxel can be fit iteratively, from which emission line parameters, including flux, velocity offset, and dispersion, can be mapped out. To obtain the structure and kinematics of the multiphase gas in the nuclear region of each galaxy, we fit the [\ion{Fe}{2}] 1.25$\mu$m, Pa-$\alpha$, and H$_2$ 2.12$\mu$m emission lines, some of the most widely detected ionized and molecular lines in the NIR. These spatial maps for each galaxy are presented in Figures \ref{fig:J1009_maps}--\ref{fig:J0906_maps}. We similarly set a S/N threshold of three and define the velocity offset, v$\rm{_{off}}$, as the offset between the emission line fit and its rest-frame wavelength. We also produced flux maps of the 3.3$\mu$m PAH emission, which are shown in Figure \ref{fig:PAH_maps}.

The spatial maps shown in Figures \ref{fig:J1009_maps}--\ref{fig:PAH_maps} were created only using a single component fit (not to be confused with the multi-component measurements used in the aperture extractions described previously and listed in Table \ref{tab:Flux_table}). While some emission lines in the nuclear spaxels are best fit with two components, mapping out these broad emission features is beyond the scope of this article. To estimate the errors of omitting these secondary components in our spatial map fits, we took a subsample of the central spaxels in each galaxy where the S/N of the broad component is the highest. We fit each of the three emission lines ([\ion{Fe}{2}] 1.25$\mu$m, Pa-$\alpha$, and H$_2$ 2.12$\mu$m) twice, first with a single component and then again with two components. When comparing the total fluxes of each method, the two component fits yielded values that were typically $\sim$10\% higher. While the inclusion of a broad, secondary component did provide a better fit, the improvements were largely seen at the wings of the emission profile. As such, differences to the primary emission profile were generally low, with velocity offset and dispersion differing less than 20 km s$^{-1}$ and $\sim$30 km s$^{-1}$, respectively. Moreover, the asymmetric profiles are mainly seen in the central, nuclear regions, and decrease as the distance increases from the nucleus.

Inspection of the nuclear spectra revealed false artifacts and distorted spectra, often labeled as `ringing' artifacts. This effect is often seen at bright, point-like sources, such as an AGN, and is present in the nuclear spaxels of following lines in our analysis: H$_2$ 2.12 $\mu$m in J0906 and J1009, and PAH in J0954 and J1009. Because this effect is caused by the undersampling of the PSF, the simplest way to mitigate these effects is to use a larger extraction aperture. While we lose the resolution of individual spaxel fits, we can still recover data over a relatively small extraction region. We first selected a box extraction covering all the affected spaxels, typically of size $0\farcs4\times0\farcs4$. The total flux, velocity offset, and dispersion were then calculated as done previously. Next, assuming the emission from the AGN can be treated as a point source, we convolved a single pixel with the NIRSpec PSF at 2.0 $\mu$m (FWHM$\sim0\farcs11$). After normalizing to the total flux measured within the box extraction, we inserted this convolved point source into the affected flux maps (indicated with an asterisk in the emission line name in Figures \ref{fig:J1009_maps}, \ref{fig:J0906_maps}, and \ref{fig:PAH_maps}). We also inserted the measured velocity offset and dispersion values into their associated maps, though it should be noted that these kinematics are unreliable on scales smaller than the extraction box.

\section{Results} \label{sec:Results}

\subsection{Aperture Extractions and Spatial Maps} \label{subsec:results_spec}

In Figure \ref{fig:spectra}, we plot the nuclear and surrounding annular extractions for each galaxy, where we have defined the nucleus as the location where the continuum emission peaks. We have also labeled all detected coronal lines along with the most commonly observed NIR emission lines. The continuum profiles in our sample are diverse, where two of the targets (J0906 and J0954) show a rising continuum towards longer wavelengths ($\alpha > 0.3$) while the other two (J0842 and J1009) show flatter continuum ($\alpha < -1.4$, see Table \ref{tab:Nuc_Properties}).

All spectra are heavily populated with a wealth of emission lines. These include luminous hydrogen recombination lines, particularly Pa-$\alpha$, Pa-$\beta$, Br-$\alpha$, and Br-$\beta$, as well as prominent low ionization lines, notably [\ion{S}{3}] 0.95$\mu$m, \ion{He}{1} 1.08$\mu$m, [\ion{Fe}{2}] 1.25$\mu$m, and [\ion{Fe}{2}] 1.64 $\mu$m. Unique to the nuclear spectra are CL emission, where the number of CLs detected varies across our sample. J0842 only has one CL detection, [\ion{Mg}{4}] 4.48$\mu$m (ionization potential, IP = 80 eV), while the other three have ten or more detections with some IPs $>$ 300 eV. Interestingly, the lowest mass galaxy in our sample, J1009, has the most CL detections with 16. Table \ref{tab:Nuc_Properties} summarizes the CL statistics of our sample (see \cite{Aravindan2026} for a more detailed analysis). Also noteworthy is the presence of PAH 3.3$\mu$m emission in our sample. With the exception of the J0906 nuclear spectra, we detect PAH emission in all of our spectra, including those featuring CLs.

In addition to the aforementioned aperture extractions, Figures \ref{fig:J1009_maps}--\ref{fig:J0906_maps} map out the flux and kinematics of prominent ionized and molecular emission lines within our coverage. The flux peaks of the different gas phases are located at the nucleus and are co-spatial with each other, including the CL emission \citep{Aravindan2026}. Beyond the nucleus, we observe diverse spatial emission structures. In some galaxies, emission is radially symmetric around the center, such as in J0842 and J0954 (with the exception of H$_2$). In others, we observe substructures, such as secondary peaks (Pa-$\alpha$ and H$_2$ in J0906, H$_2$ in J0954, and Pa-$\alpha$ in J1009) and asymmetric, extended emission (H$_2$ in J0906 and J1009). 

The kinematics of the gas are also varied in all phases but generally do not exceed projected velocities of $\pm$100 km s$^{-1}$. Only J0906 shows a clear gradient between the blueshifted and redshifted gas that cuts across the galaxy nucleus, which is most pronounced in Pa-$\alpha$. For J0842 and J1009, the gas are mostly blueshifted relative to the rest-frame of the galaxy. Inspection of the dispersion maps yield widths of 50 -- 75 km s$^{-1}$ in all galaxies. Although certain regions do show dispersion as high as 200 km s$^{-1}$, including the nucleus of J0906 and some substructure in J0842, we find no clear spatial variations in the rest of the extended emission. Further details and interpretation of these kinematic maps will be discussed in Section \ref{subsec:Discussion_maps}.

The PAH flux maps of Figure \ref{fig:PAH_maps} are likewise varied in structure, with both radially symmetric emission around the nucleus (J0842 and J0954) and complex substructures (J0906 and J1009) observed. Notably, there is a distinct `cavity' in the PAH flux map of J0906, where no emission is detected in the central $\sim$300 pc. This is consistent with other Seyfert nuclei which often emit sufficiently powerful hard photon fluxes to destroy small PAH grains \citep{Lai2023}. The nuclear spectrum of J0906 (fourth row of Figure \ref{fig:spectra}) also shows this lack of emission. In addition, the PAH emission in J1009 extends to the NE and SW, which matches the orientation of the 2.0$\mu$m continuum image. This, however, is contrasted by the H$_2$ emission which extends perpendicularly to the NW direction.

\begin{deluxetable}{ccccccc}
\centering
\setlength{\tabcolsep}{8pt}
\caption{Galaxy and Nuclear Properties} 
\label{tab:Nuc_Properties}
\tablehead{\colhead{Galaxy} & \colhead{Log($M_{*}/M_\odot$)} & \colhead{log(dE$\rm{_{out}}$/dt)} & \colhead{Spectral Index, $\alpha$} & \colhead{CL $\#$ (highest IP)} & \colhead{log(N$\rm{_{H_2, Total}}$)} & \colhead{M$\rm{_{H_2}}$}\\
\colhead{} & \colhead{} & \colhead{(erg s$^{-1}$)} & \colhead{} & \colhead{} & \colhead{(cm$^{-2}$)} & \colhead{(M$_\odot$)}}
\startdata
J1009 & 8.77 & 37.5$^{+0.3}_{-0.3}$ & -1.99, -1.42 & 16 (447 eV) & 15.29 (15.38*) & 0.8 (1.0*)\\
J0954 & 9.12 & 38.7$^{+0.3}_{-0.3}$ & -0.96, 0.39 & 14 (351 eV) & 15.78 & 13.4\\
J0842 & 9.34 & --- & -2.27, -2.50 & 1 (80 eV) & 15.88 & 13.1\\
J0906 & 9.36 & 38.6$^{+0.5}_{-0.5}$ & -0.04, 0.38 & 11 (447 eV) & 15.61 & 18.6
\enddata
\tablecomments{Columns: (1) Galaxy, in order of increasing galaxy stellar mass. (2) Galaxy stellar mass from the NSA catalog. (3) Kinetic energy outflow rate based on [\ion{Si}{6}] from \cite{Aravindan2026}. (4) Continuum spectral index of the nuclear spectrum. The first value is measured from 1.0 - 3.0 $\mu$m, while the latter is 3.0 - 5.0 $\mu$m. Average errors are $\sim$0.05. (5) Number of coronal lines detected. The value in the parenthesis refers to the highest potential energy of the line detected. (6-7) Hot molecular H$_2$ gas column density and mass (see Section \ref{subsec:results_H2}). The quoted values do not include a heavy element correction. The average errors are around 25$\%$.}
\tablenotetext{*}{H$_2$ gas column density and mass of the NW region of the plume (see dotted cyan circle of Figure \ref{fig:J1009_maps}).}
\end{deluxetable}

\subsection{Continuum Emission} \label{subsec:continuum_results}

As mentioned in Section \ref{sec:Analysis}, we fit the nuclear continuum with \textsc{CAFE} in order to determine the relative contributions of the stellar, dust, and AGN contributions. However, due to the lack of MIR data to anchor the modeling, we avoid making any detailed conclusions based on a single fit. Instead, we highlight the general results and trends that consistently emerge across multiple fitting iterations. For J0842, the continuum emission is largely dominated by the stellar component, all except for a small contribution from dust at the reddest wavelengths. The continuum for both J0906 and J0954, however, are characterized with an increasing flux towards longer wavelengths. While the stellar component is still dominant at bluer wavelengths, our modeling cannot replicate the rising continuum without significant contribution from an AGN, either in the form of an accretion disk or hot dust. Lastly, model fits to J1009 indicate a stronger contribution from the stellar component than J0906 and J0954. However, like these two, the stellar model alone cannot replicate the full continuum emission without some contribution from either an accretion disk or hot dust. A sample set of continuum fits using \textsc{CAFE} are provided in Figure \ref{fig:CAFE_fits} of the Appendix.

\subsection{Masses of the Hot H$_2$ Gas} \label{subsec:results_H2}

Across our NIRSpec datasets, we detect 36 unique rotational and rovibrational H$_2$ emission lines at $>$2.5$\sigma$ level, with excitation energies, $E_u/k$, up to 21,400~K. This large number of excitation states allows us to directly obtain gas masses and temperatures of the hot H$_2$ component. We followed the prescriptions described in \cite{Youngblood2018} to construct H$_2$ excitation diagrams from which temperatures and masses can be calculated. First, we took fluxes from Table \ref{tab:Flux_table} and corrected them for extinction. Extinction values were taken from the Pa-$\alpha$/Pa-$\beta$ ratios of Table 3 of the companion paper \cite{Aravindan2026}, where E(B-V) values range between 0.00 and 0.44. As a check, we compared these values to other extinction estimates based on strong Brackett and Pfund lines, and confirmed that they are all consistent within the uncertainties. Before applying our temperature model, we incorporated the magnetic dipole term in the Einstein A-coefficient presented in \cite{Roueff2019}. We used a single temperature model, where the slope and y-intercept were kept as free variables. This linear model follows Equation (2) from \cite{Youngblood2018}, where the slope is used to estimate the gas temperature. By imputing this temperature into the partition function \textit{Z(T)}, the total column density of the H$_2$ gas can be calculated, from which the gas mass can be derived. Here, we assumed an ortho-to-para ratio of 3.

In Figure \ref{fig:H2_excite}, we plot H$_2$ excitation diagrams which show the linear fits to the data, based on the nuclear H$_2$ fluxes. For the purposes of this article, we only fit the $v$ = 0--0 and $v$ = 1--0 transitions, about 15 lines for most targets, with the linear model to avoid the complexity of fitting the higher rovibrational transitions which do not appear to follow a linear (nor power-law) trend with the rotational lines. The resulting temperatures are around 3,000 K with the only exception being J0906 which has a calculated temperature of just under 6,000 K. This value however, was only calculated from three data points (see lower-right panel of Figure \ref{fig:H2_excite}). The measured column densities are more consistent, averaging log(N$\rm{_{H_2}}$/cm$^{2}$) $\sim$ 15.5 across all galaxies. From the derived temperatures and column densities, the estimated hot H$_2$ gas masses for our sample (uncorrected for He and other heavy elements) range from 1 -- 19 M$_\odot$ (see Table \ref{tab:Nuc_Properties}). We note that J1009 differs by an order of magnitude and this could be due to the smaller physical size of the aperture extraction used (see Section \ref{subsec:Discussion_H2} for further discussion).


\section{Discussion} \label{sec:Discussion}

The improved spectral resolution and sensitivity of \textit{JWST} over previous facilities has uncovered diverse nuclear conditions in our sample. As shown in Figures \ref{fig:spectra} -- \ref{fig:PAH_maps}, each galaxy shows varied spectral features and spatial emission line structures. The extended wavelength coverage from 0.9 -- 5.1$\mu$m has also enabled us to detect over 100 emission lines at the 3$\sigma$ level, a significant increase compared to previous ground-based observations. In the following sections, we will discuss the nuclear spectra and the multi-phase gas structure of these unique systems.

\subsection{Nuclear Spectra} \label{subsec:Discussion_spec}

\subsubsection{Continuum and Emission Line Characteristics} \label{subsubsec:continum_emission_lines}

As described in Section \ref{subsec:continuum_results}, continuum modeling the sample reveals a significant contribution from the 10 and 100 Myr starburst templates in the J and H bands. The contribution in K band and redwards, however, varies greatly in our sample. For J0842, much of the continuum within our coverage can be modeled using stellar templates. In contrast, the continua of J0906, J0954, and J1009 cannot be described by stellar emission alone. While the relative contributions between stellar and hot dust in J1009 are unclear, J0906 and J0954 show a clear rising continuum, a characteristic of emission from hot dust or an AGN accretion disk. Here, J0906 is noteworthy since only the most prominent emission lines are detected in its nuclear spectrum at redder wavelengths, a stark contrast to the rest of the sample. A contributing cause to this could be dilution due to a strong continuum \citep{Rodriguez2011}. Loss of line contrast, perhaps due to significant emission from an AGN or hot dust, would cause weaker emission lines to fall below our detection limit. 

As noted earlier, our coverage includes important diagnostic lines, whose line ratios can be used to identify dominant ionizing or heating sources. Based on the [\ion{Fe}{2}]/Pa$\beta$ and 1--0 S(1) H$_2$/Br$\gamma$ diagnostic, we find that the nuclear flux ratios are consistent with those of other Seyferts \citep{Riffel2013, Colina2015}. Additionally, all targets in this study have at least one CL detected in their nuclear spectrum. This is most evident in J0906, J0954, and J1009 where more than ten CLs are detected in each galaxy, with some having IPs $>$300 eV. Coupled with the rising continuum slope, these all indicate that the AGN are a primary energizing source within the central regions of these systems.

\begin{figure}
\centering
\epsscale{1.2}
\plotone{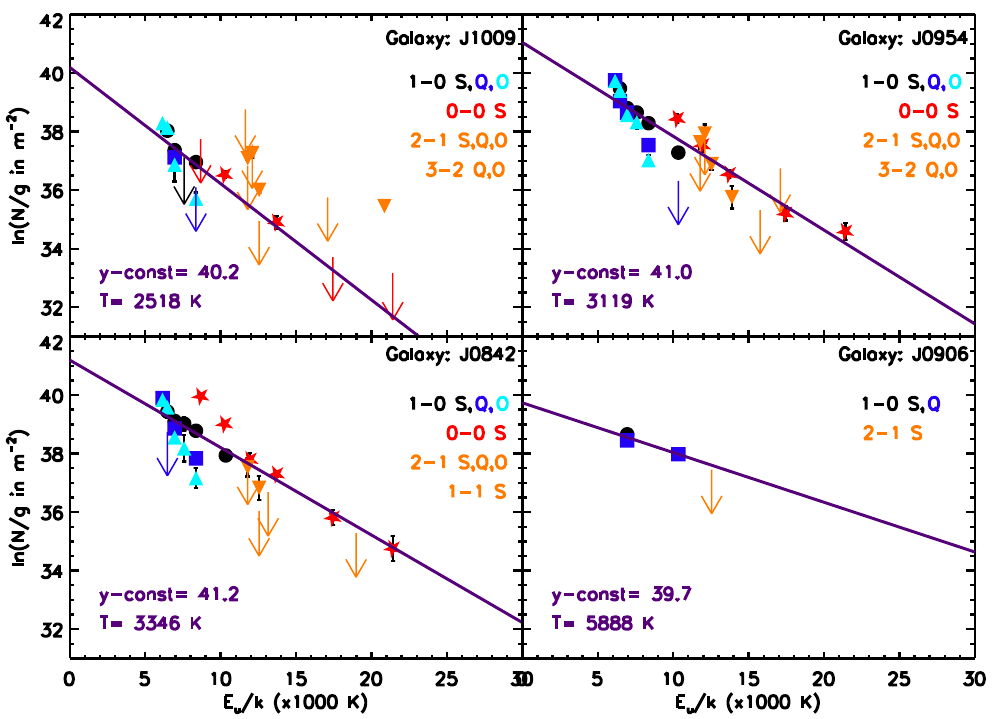}
\caption{H$_2$ excitation diagrams for the four galaxies in our sample. The 0--0 and 1--0 transitions are labeled separately while the higher transitions are grouped together. Upper limits are shown as downward arrows (see Table \ref{tab:Flux_table}). The linear model is fit only to the 0--0 and 1--0 transitions, and the resulting temperature and y-constant (intercept) are listed. \label{fig:H2_excite}}
\end{figure}


The main exception to this, however, is J0842 where only one CL, [\ion{Mg}{4}] (IP = 80 eV), is detected in the nucleus. Moreover, the detection of [\ion{Mg}{4}] alone can be plausibly explained by shocks and other stellar-driven feedback \citep{Pereira2024b}. To further assess the nuclear source in J0842, we invoke additional line diagnostics from \cite{Costa2026}. Based on their 1--0 O(5) H$_2$/PAH vs. 1--0 S(1) H$_2$/Br$\gamma$ diagram (in addition to [\ion{Fe}{2}]/Pa$\beta$ vs H$_2$/Br$\gamma$), we find the nuclear emission of J0842 to be consistent with an AGN. It should be noted, however, that these diagnostics can be affected by projection effects (the molecular and ionized species may not necessarily reside in the same regions) and that they are based only on photoionization modeling, without shocks included.

In addition to these diagnostics, off-center [\ion{Ar}{6}] (IP = 75 eV) emission is detected $\sim$150 pc SW of the J0842 nucleus, along with [\ion{Mg}{4}] (see Figure \ref{fig:J0842_ArVI}). \cite{Pereira2024b} report the detections of both [\ion{Mg}{4}] and [\ion{Ar}{6}] as an indication of AGN activity. However, the off-center nature in J0842 is particularly compelling. Further discussion of this can be found in the Appendix.

The lack of higher ionization CL emission does not necessarily rule out an AGN. Previous studies have found that only 40$\%$ of Seyferts have CL detections, with the detection rate falling to $\sim$20$\%$ in Seyfert 2 galaxies \citep{Lamperti2017}. Indeed, well known Seyferts such as Mrk 231 have no detected CLs in the IR \citep{Inami2013,Stierwalt2014,AlonsoHerrero2024}. All put together, the evidence presented above suggest J0842 houses a less luminous AGN, perhaps one that is masked or diluted by significant stellar emission. This narrative is consistent with its location in the composite region of the BPT and, as described in Section \ref{subsec:continuum_results}, the large stellar contribution to the nuclear continuum.

\subsubsection{Black Hole Mass}

\begin{deluxetable}{ccccccc}
\setlength{\tabcolsep}{7pt}
\caption{AGN Radiation Field and Gas Morphology Summary} 
\label{tab:AGN_PAH_prop}
\tablehead{\colhead{Galaxy} & \colhead{EQW$\rm{_{PAH, 3.3}}$ Nuc. (Ann.)} & \colhead{Log(L$\rm{_{Bol}}$)} & \colhead{$\rm\frac{[Mg\;VIII]}{[Mg\;IV]}$} & \colhead{Ionized Morph.} & \colhead{H$_2$ Morph.} & \colhead{PAH Morph.}\\
\colhead{} & \colhead{(\AA)} & \colhead{(erg s$^{-1}$)} & \colhead{} & \colhead{} & \colhead{} & \colhead{}}
\startdata
J1009 & 177$\pm$7 (341$\pm$8) & 41.3--42.8 & 2.5 & clumpy & plume, clumpy & elongated\\
J0954 & 144$\pm$15 (290$\pm$13) & 42.3--44.1 & 1.2 & symmetric & off-center clump & symmetric\\
J0842 & 205$\pm$9 (244$\pm$11) & 42.3--43.5 & --- & symmetric & symmetric & symmetric\\
J0906 & 0 (100$\pm$17) & 42.6--43.9 & 1.1 & clumpy & extended, clumpy & center suppressed
\enddata
\tablecomments{Columns: (1) Galaxy, in order of increasing stellar mass. (2) PAH 3.3$\mu$m equivalent widths using the nuclear and annular extractions. (3) Range of AGN bolometric luminosities, derived from relations from \cite{Kim2022} and \cite{Spinoglio2024}. Intrinsic scatter in these relations are $\sim$0.3 dex. (4) Flux ratio of [\ion{Mg}{8}] (225 eV) / [\ion{Mg}{4} (80 eV)]. (5--7) Morphologies of the different gas phases based on the flux maps of Figures \ref{fig:J1009_maps}--\ref{fig:PAH_maps}. `Symmetric' indicates the distribution of the emission is radially symmetric around the nucleus, and `clumpy' signifies that the emission is not smoothly distributed.}
\end{deluxetable}



\begin{figure}
\centering
\epsscale{1.2}
\plotone{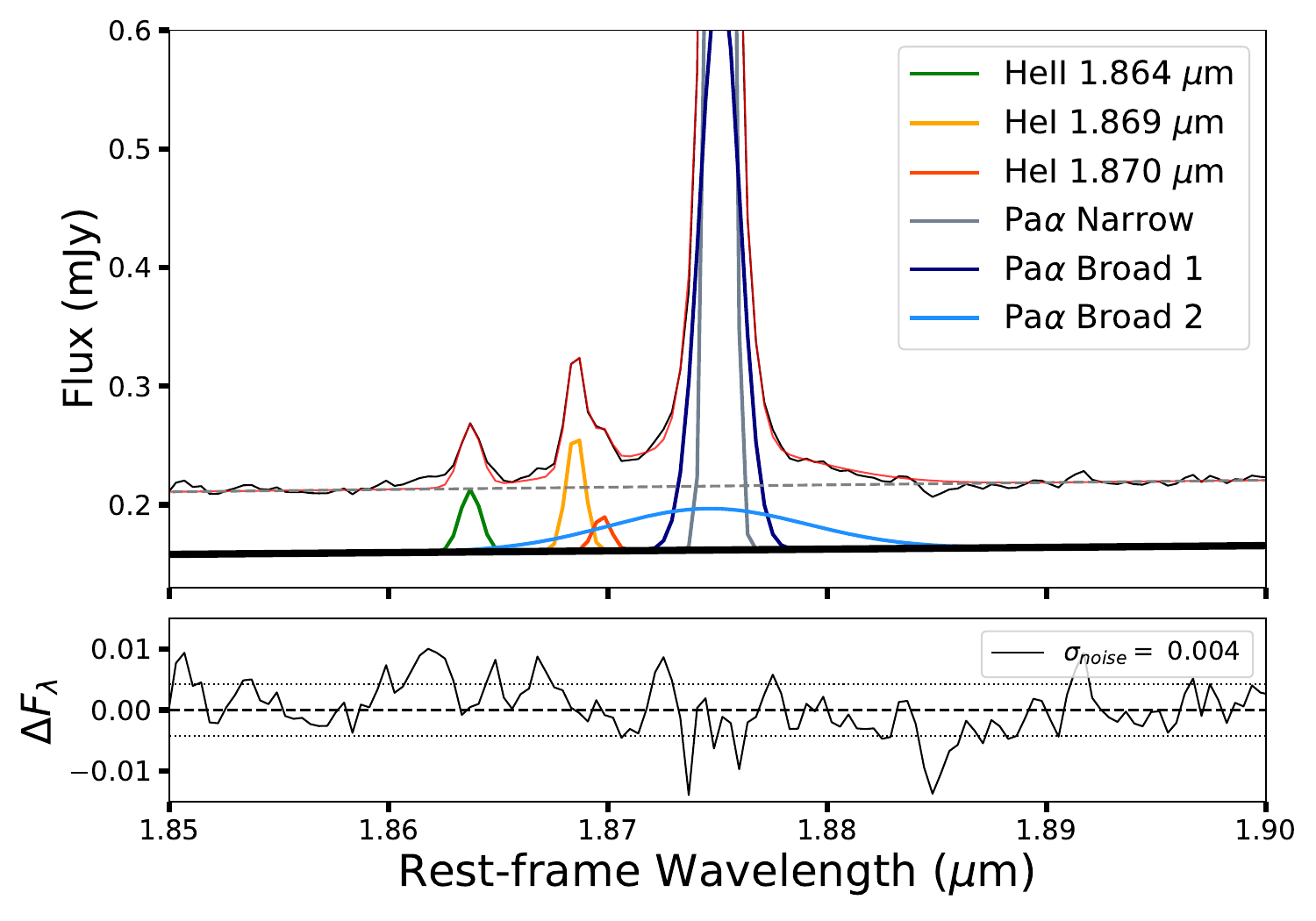}
\caption{MCMC fit of the Pa$\alpha$ emission complex in J0954. Three Gaussian components are fit to Pa$\alpha$, where we assume the broadest (`Broad 2') is originating from the broadline region. Following virial mass estimators, we calculate a M$\rm{_{BH}}$ = 10$^{6.01\pm0.21}$ M$_{\odot}$, placing it within the scatter of the local M$\rm{_{BH}}$ - M$\rm{_{*}}$ relation. \label{fig:J0954_Pa}}
\end{figure}

Previous studies have reported black hole masses for our targets through either virial mass estimates from broad H$\alpha$ emission or Eddington luminosities \citep{Reines2013,Moran2014,Marleau2017,Manzano2019}. The masses reported broadly range from 10$^{5.0-6.5}$ M$_{\odot}$, and fall within the scatter of the local M$\rm{_{BH}}$ - M$\rm{_{*}}$ relation \citep{Reines2015,Bohn2020}. While we detect broad components in many of our hydrogen lines, it is difficult to distinguish whether these features are from outflowing gas or the broadline region. Only one target, J0954, required three gaussian components for Pa$\alpha$, the broadest of which has a FWHM of 1,700 km s$^{-1}$ (see Figure \ref{fig:J0954_Pa}), typical of broadline profiles \citep{Reines2015,Lamperti2017}. If we assume this third, most broad emission component originates from the broadline region, we can estimate a virial mass. Following equation 10 from \citet{Kim2018}, who adopted a virial factor of log $\mathnormal{f}$ = 0.05 $\pm$ 0.12 \citep{Woo2015}, we calculate the BH mass as follows,

\begin{equation}
\label{eq:1}
\small \frac{M_{BH}}{M_{\odot}} = 10^{7.07\pm0.04}\left(\frac{L_{\rm{Pa}\alpha}}{10^{42} \;\rm{erg}\;\rm{s}^{-1}}\right)^{0.49\pm0.06}\left(\frac{\rm{FWHM}_{\rm{Pa}\alpha}}{10^3\;\rm{km}\;\rm{s}^{-1}}\right)^2
\end{equation}

where L$_{\rm{Pa}\alpha}$ was derived from the flux of the third Pa$\alpha$ component, as listed in Table \ref{tab:Flux_table}. This results in a BH mass of M$\rm{_{BH}}$ = 10$^{6.01\pm0.21}$ M$_{\odot}$, which lies in between previous measurements of 10$^{5.0}$ M$_{\odot}$ \citep{Reines2015} and 10$^{6.46}$ M$_{\odot}$ \citep{Marleau2017}, and is consistent with other BH masses found within local galaxies with similar stellar masses. However, as mentioned previously, it is difficult to disentangle and identify the various emission sources in the Pa$\alpha$ complex (see Figure \ref{fig:J0954_Pa}). Thus, we caution that the broadline component used may not be tracing the gas in the broadline region.  

\subsubsection{The 3.3$\mu$m PAH emission} \label{subsubsec:33PAH}




Polycyclic aromatic hydrocarbon (PAH) emission can serve as a sensitive indicator to the radiative state of the local ISM. This is because strong radiation, such as that produced by an AGN, often destroys the relatively fragile PAH grains \citep{Diamond2010,Lai2023}. Indeed, observations isolating the central sub-kpc emission of active galaxies reveal strong suppression and often complete lack of PAH emission \citep[e.g.,][]{Honig2010,Armus2023}. We observe a similar scenario in J0906, where no PAH emission is detected within a `cavity' of $\sim$300 pc around the nucleus. Interestingly, \cite{Yang2020} report enhanced radio emission indicative of a jet within the nuclear regions, however its scale ($\sim$50 pc, projected) is smaller than the size of cavity. While jets have been shown to suppress PAH emission \citep{Ogle2010}, the effects here may only be marginal due to the small scales of the detected emission.

The nuclear spectra of the other targets show a different scenario. Clear PAH emission is detected in both nuclear and off-nuclear apertures, with emission seen within $\sim$100 pc of the nuclei. While PAH emission is generally observed to be suppressed near AGN, some studies have instead found PAH excitation \citep{Smith2007,Jensen2017}. Our sample presents a rather interesting case since we detect the smaller grains that are more susceptible to strong radiation fields, and thus are expected to be more easily destroyed by an AGN. Without MIR spectra, however, it is impossible to explore this further, such as determining the size distribution of the PAHs and the mechanisms governing their excitation. Still, we can make a speculative assessment as to the factors that could explain the relative abundance of PAH emission detected so close to the AGN. 

First, the centralized PAH emission could be due to a compact region of star-formation \citep{DiazSantos2010a}. This could be the case for J0842 and J1009 since their nuclear continua are best fit with a strong contribution from starburst templates. J0842 is also located in the composite region of the BPT, suggesting a mix of AGN and stellar activity. Second, the radiation from the AGN may not be sufficient to destroy all the PAH grains. Or alternatively, instead of destroying the PAH grains, the central AGN could be acting as an energizing source to induce PAH emission. \cite{Jensen2017} were able to replicate spatially resolved PAH emission in a sample of nearby Seyfert galaxies using \textsc{CLOUDY} modeling, where the AGN were shown to provide the required energy for PAH excitation out to $\sim$500 pc. To provide enough self-shielding to allow the PAH grains to survive, they assumed a dense column density of $N_{\rm{H}}$ = 10$^{23}$ cm$^{-2}$. It should be noted however that while this dense medium may be able to prevent the PAH grains from being destroyed, their emission may also be obscured from our line of sight. Still, if the AGN is indeed contributing to the PAH emission, then any star formation estimated based on the PAH emission from the central region in these targets could be significantly overestimated .

\subsection{The AGN Radiation Field} \label{subsec:AGN_radiation}

We next assess the radiation field from the AGN and its impact on the surrounding ISM. PAHs are a common tracer of the ionization state of the ISM due to their sensitivity to harsh radiation environments. In Table \ref{tab:AGN_PAH_prop}, we list the PAH equivalent widths (EQW) measured from the nuclear and annular extractions of each galaxy. Comparing the two regions, we see higher EQWs in the outer, annular regions, suggesting some suppression of the PAH emission in the nuclear regions. There are several possible explanations for this. As mentioned previously, this deficit can stem from the ionization or destruction of the PAH molecules. Alternatively, AGN can heat the nearby dust, causing an increase in the continuum level which can lead to the dilution of the PAH emission. \cite{Desai2007} note a trend of decreasing PAH EQW with increasing 24 $\mu$m luminosity, which samples the peak emission of the warm ($\sim$100 K) dust emission. We see a similar trend in our dataset where J0906 and J0954, the two targets with the highest spectral index $\alpha$ and lowest PAH 3.3 $\mu$m EQW, are the most luminous in the 22 $\mu$m WISE W4 band. As such, for these two targets, the effects of continuum dilution may be stronger in suppressing the PAH emission. However, without observations of the larger PAH grains at MIR wavelengths, it is difficult to determine which process is responsible for the overall PAH deficit.

We further assess the relative strength of each radiation field in our sample by examining the emission line ratios of like species in different ionization states. The ratios of [\ion{Si}{6}] 167 eV with [\ion{Si}{9}] 304 eV and [\ion{Si}{10}] 351 eV are an excellent choice for this analysis. Being transitions of the same species, these ratios are independent of relative abundance and metallicity effects, reducing the overall degeneracy in determining the ionization state of the gas. From the fluxes listed in Table \ref{tab:Flux_table}, both [\ion{Si}{9}] / [\ion{Si}{6}] and [\ion{Si}{10}] / [\ion{Si}{6}] have ratios between 0.20 and 0.30 for J0906 and J0954. For J1009, however, the [\ion{Si}{9}] / [\ion{Si}{6}] ratio is comparatively higher at 0.58. A similar trend is seen in [\ion{Mg}{8}] 225 eV / [\ion{Mg}{4}] 80 eV where the ratio for J1009 is two times higher than J0906 and J0954 (see Table \ref{tab:AGN_PAH_prop}). While [\ion{Si}{10}] falls within the detector gap for J1009, these measured ratios indicate that J1009 has a harder radiation field, and/or more gas in higher ionization states. One possible cause for this could stem from the lower mass BH that is expected to reside in J1009 (assuming the $M_{\rm{BH}}$ -- $M_{*}$). Through theoretical modeling, \cite{Sun1989} and \cite{Cann2018} show that the hardening of the spectral energy distribution increases with decreasing BH mass for a given Eddington ratio. This is because the Schwarzschild radius of a less massive BH is smaller, leading to higher accretion disk temperatures where higher ionization lines are more easily produced. 

When we compare the relative emission line strengths of CLs of different ionization potentials to those observed in other Seyfert galaxies, we find that those in J1009 are the most similar to what is observed in well-studied Seyfert 1 and Seyfert 2 galaxies, such as NGC 1068 \citep{Spinoglio2005}, NGC 4151 \citep{Sturm1999}, and the KONA sample \citep{Muller2018b}. In contrast, the relatively strong [\ion{Mg}{4}] line in J0906 compared to its undetectably weak [\ion{Si}{9}] and [\ion{Ca}{8}] emission makes J0906 different from many of the previously studied narrow line regions of Seyfert galaxies. 

While it is difficult to estimate the AGN bolometric luminosities of our sample due to the lack of direct measurements of the AGN continuum at high frequencies, we can still make indirect estimates. To do this, we follow the prescriptions of \cite{Spinoglio2024} and \cite{Kim2022}, where the former uses K-band luminosity while the latter uses hydrogen recombination lines. We first calculate the K-band luminosities, L$_{2.2 \mu m}$, from our nuclear extractions. We input these luminosities into the L$_{2.2 \mu m}$ -- L$\rm{_{Bol}}$ relation derived by \cite{Spinoglio2024}, where we obtain log(L$\rm{_{Bol}}$/erg s$^{-1}$) of 2E+41, 2E+42, 2E+42, and 4E+42 for J1009, J0954, J0842, and J0906, respectively. Alternatively, \cite{Kim2022} use the luminosities of Pa$\alpha$, Pa$\beta$, Br$\alpha$, and Br$\beta$ to calculate L$\rm{_{Bol}}$. The derived bolometric luminosities from each line (see Table \ref{tab:Flux_table}) are consistent with each other but are generally over a magnitude higher than those from L$_{2.2 \mu m}$. We note, however, that the scatter within each relation is large ($\sim$0.3 dex). Also, the galaxies used in these studies are more massive with more luminous AGN than the dwarf sample used here. As such, the relations presented above may not be well-constrained at lower masses. Still, we list the range of bolometric luminosities in Table \ref{tab:AGN_PAH_prop}, but due to the uncertainties involved, we only treat them as coarse estimates.

Lastly, the diverse continuum slopes of our sample, specifically at wavelengths 3-5 $\mu$m, suggest varying degrees of hot dust contribution. Inspection of the slopes, however, does not lead to any clear correlation with the hardness of the radiation field, as measured by the available CL ratios.

\subsection{The Hot H$_2$ Gas Component} \label{subsec:Discussion_H2}


In Section \ref{subsec:results_H2}, we analyzed the hot H$_2$ gas component from which temperatures, column densities, and masses were derived using the excitation diagrams presented in Figure \ref{fig:H2_excite}. The resulting values were generally around 3,000 K and log(N$\rm{_{H_2}}$/cm$^{2}$) $\sim$ 15.5, with gas masses on the order of tens of M$_\odot$ (see Table \ref{tab:Nuc_Properties}). There are two notable exceptions that differ from the rest of the sample. The first is J0906, which has an estimated gas temperature of roughly 6,000 K. Although the fit is only based on three data points (1--0 S(1), 1--0 Q(3), and 1--0 Q(7)) and thus quite uncertain, it is still likely that the temperature of the hot gas is comparatively high. One reason for this is that the relative flux differences between these values are lower in J0906 when compared to the rest of the sample. This causes the slope of the fit (slope = -1/T) to be more shallow, leading to higher temperatures. While a temperature of 6,000 K may be unphysical, recent work has found gas temperatures of over 5,000 K in regions of shocked gas \citep{Roueff2023}. As mentioned previously, J0906 is the only target with evidence of a radio jet \cite{Yang2020} where shock heating could exist. However, regardless of the mechanism, it is likely that only a small fraction of the gas is heated to such temperatures. As such, we caution taking the reported temperatures for J0906 at face value. The second notable exception is J1009, where the estimated hot H$_2$ mass is a factor of 10 lower than the rest of the sample. This, however, is likely due to its closer proximity ($\sim$60 Mpc) compared to the rest of the sample ($\sim$130 -- 210 Mpc), and thus we are sampling a smaller region by an order of magnitude.

To facilitate comparisons with our sample, we draw from other hot molecular gas studies of nearby galaxies from the literature \citep{Lopez2025,Doan2025}. Using their line fluxes of nuclear regions, we recalculate the gas temperatures, column densities, and masses using the methods described in Section \ref{subsec:results_H2}. For simplicity, we only use the $v$ = 1--0 transitions. We find that our sample has as much as an order of magnitude lower column densities when compared to the higher mass systems, such as the Seyfert galaxy M58 \citep{Lopez2025}. However, the gas masses of our sample are slightly below that of M58 (19 M$_\odot$) and above the low metallicity dwarf (0.7 M$_\odot$) presented in \cite{Doan2025}. It should be noted that the circular extraction sizes of these studies, r$\sim$50 pc, are smaller than those used here, r$\sim$100--250 pc. If we were to reduce our aperture size, we can expect the observed molecular column density to increase given that the main excitation source is centrally located (a reasonable assumption given the ionized emission peaks at the nuclei). While the relative extraction size can affect the measured column densities and gas masses, we find that the dwarf galaxies in this sample can heat and excite the molecular gas almost to the same degree as more massive galaxies. This is particularly interesting considering that the M$\rm{_{BH}}$ of these dwarfs are expected to be roughly two orders of magnitude lower.


Closer inspection of the excitation diagrams of Figure \ref{fig:H2_excite} show an alignment between the pure rotational lines and the lower ($v$ = 1--0) rovibrational transitions, where a single temperature model can sufficiently characterize the state of the molecular gas. However, there is also a decoupling seen with the higher rovibrational transitions which lie above the rotational curve (see J1009 in Figure \ref{fig:H2_excite}). This decoupling with the higher rovibrational transitions has been noted in other systems, such as the infrared luminous radio galaxy 4C12.50 \citep{Villar2023}. In contrast, \cite{Lopez2025} report a decoupling of the rotational and $v$ = 1--0 rovibrational transitions in M58. While the former reports heating induced by shocks from a radio jet, the latter mainly attributes the discrepancy due to sub-thermal excitation in low density regions ($\rm{n_H}<n_{\rm{crit}}$, $\sim$10$^{6}$ cm$^{-3}$ for rovibrational transitions). However, since the low-J pure rotational lines are not covered by our observations, we cannot determine the origin of the rovibrational decoupling in our sample.



\begin{figure}
\centering
\epsscale{1.2}
\plotone{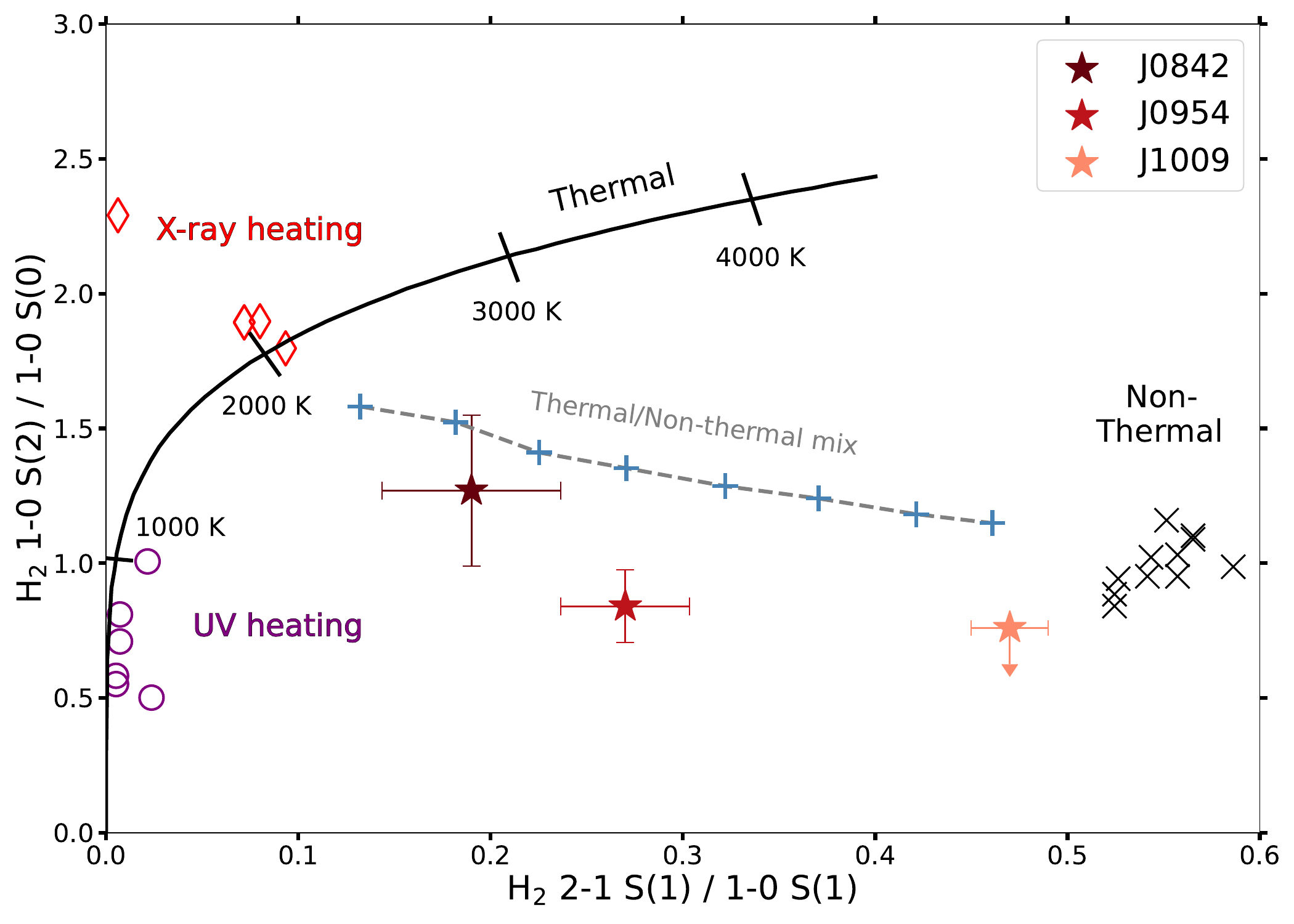}
\caption{1--0 S(2) 2.03 $\mu$m/1--0 S(0) 2.22 $\mu$m vs. 2--1 S(1) 2.25 $\mu$m/1--0 S(1) 2.12 $\mu$m emission line ratios of our sample. J0906 is omitted since the required emission lines are not detected. The solid curve indicates pure thermal emission where temperatures at every 1,000 K are marked \citep{Mouri1994}. Ratios based on modeling from X-ray heating \citep{Drain1990}, UV heating \citep{Sternberg1989}, and non-thermal UV fluorescence \citep{Black1987} excitation mechanisms are also plotted. The dotted gray line traces ratios based on a mixture of thermal and non-thermal processes. Ratios start at a thermal -- non-thermal ratio of 90--10 (\textit{left}) and end at 20-80 (\textit{right}), where each step represents a shift of 10$\%$. The ratios of our sample indicate contribution from both thermal and non-thermal processes. \label{fig:H2_thermal_nontherm}}
\end{figure}

It is thus clear that multiple excitation processes are heating the gas in our sample. Ratios of H$_2$ lines that are sensitive to different excitation mechanisms (i.e. thermal or non-thermal processes) can be used to determine which is dominant. Non-thermal processes, such as fluorescence, are more efficient at populating the higher vibrational levels ($v\geq2$) and thus lines such as $v$ = 2--1 S(1) can be used to identify non-thermal excited regions. Following the modeling and line ratios of \cite{Mouri1994}, we use the 2--1 S(1) 2.25 $\mu$m/1--0 S(1) 2.12 $\mu$m ratio, where the latter is readily detectable and has excitation that is largely thermal. Using the available fluxes listed in Table \ref{tab:Flux_table} (note that some values for J0906 and J1009 are either upper limits or not detected), these ratios confirm a mix of thermal and non-thermal contributions in all four galaxies. As shown in Figure \ref{fig:H2_thermal_nontherm}, the relative contributions differ with each galaxy. For instance, J0842 has the lowest 2--1 S(1) / 1--0 S(1) ratio and its proximity to the thermal model track suggests a 80$\%$--20$\%$ contribution from thermal and non-thermal process, respectively. On the other hand, this ratio is higher in J1009, an indication of stronger non-thermal contributions. This is in accord with its lower temperature estimates, suggesting that thermal processes are not as dominant. These trends are reflected in the excitation diagrams of Figure \ref{fig:H2_excite}, where the higher transition rovibrational lines ($\nu$ = 3$\rightarrow$2, $\nu$ = 2$\rightarrow$1, and $\nu$ = 1$\rightarrow$1) are enhanced in J1009. This stark difference in the H$_2$ excitation could be connected to the AGN radiation fields. As discussed previously in Section \ref{subsec:AGN_radiation}, the harder radiation field of J1009 could be driving the stronger non-thermal excitation in the H$_2$ gas, resulting in higher rovibrational transitions to be more prevalent. To confirm this trend however, further observations providing both measurements of the purely rotational transitions and more substantial statistics will be necessary.

\subsection{Global Star Formation Rates}

Star formation rates (SFRs) are an important metric in evaluating AGN feedback but can be difficult to measure due to the AGN acting as an additional energizing source. A number of star formation estimators are available, though each has its own set of caveats. As such, our aim is only to provide a global estimate to the SFRs within our FoV, and verify if they fall on the main sequence (MS) of star forming galaxies \citep{Renzini2015,McGaugh2017}. PAH emission has been a popular MIR star formation indicator, and recent effort has been made to formulate a calibration using the 3.3$\mu$m PAH \citep{Lai2020,Gregg2025}. To estimate the total SFR of each galaxy, we integrated over all the 3.3$\mu$m PAH emission within our full FoV. We then follow Equation (1) of \cite{Lai2020}, which was calibrated through the [\ion{Ne}{2}] 12.8 $\mu$m and [\ion{Ne}{3}] 15.6 $\mu$m SFR indicator. The resulting SFRs in our sample range between 0.05 -- 0.29 M$_\odot$ yr$^{-1}$, placing them within 0.3 dex of the MS.

We next explore H$\alpha$, one of the most common methods of estimating SFRs. While this has been routinely used for star forming galaxies, it is well known that it can be compromised in AGN systems. In an effort to separate the stellar and AGN contributions to the H$\alpha$ emission, \cite{Malkan2025} create a `mixture' diagram using a large sample of Type I and II Seyferts from SDSS. By evaluating their BPT line ratios, and thus their positions on the diagram, they estimate the relative contributions to H$\alpha$ from the AGN and stars. Here, we opted to use the SDSS data of our sample in order to best match their diagnostic. Since the SDSS line ratios of our sample do not fall directly on the model tracks of \cite{Malkan2025}, we trace a curve modeled after the \cite{Kewley2001} diagnostic line. Following this, the contributions to H$\alpha$ from stellar processes range from 49$\%$ (J1009) to 72$\%$ (J0842). We then multiply the SDSS fluxes by these percentages and calculate the SFRs using the \cite{Kennicutt1994} model,

\begin{equation}
\label{eq:2}
\rm{SFR} = \frac{L_{\rm{H}\alpha}}{1.26\times10^{41}erg \;\rm{s^{-1}}} M_\odot\;yr^{-1}.
\end{equation}

The resulting SFRs, 0.027 -- 0.26 M$_\odot$ yr$^{-1}$, are consistent with those values based on the PAHs.

The above values and their placement on the MS relation indicate that feedback processes have yet to quench the overall star formation of the galaxies. Instead, it is likely that the AGN is only having a local effect, perhaps within the central hundreds of parsecs. Estimating the SFR in close proximity to the AGN, however, is difficult due to the high degree uncertainty in the available diagnostics. Moreover, considering the upper limit nature of these calculations along with the presence of outflows, it cannot be ruled out that our sample could be on a transitional phase where the feedback effects are just starting to take place. This is particularly true for J1009 which falls close to 0.3 dex below the MS.

\subsection{Multi-phase Gas Structure} \label{subsec:Discussion_maps}

\subsubsection{Nuclear and Spatially-resolved Emission} \label{subsubsec:substructure}

The flux maps of Figures \ref{fig:J1009_maps}--\ref{fig:PAH_maps} show the most luminous emission of the different gas phases are generally concentrated in the nuclear region and are co-spatial with each other. While the structure of the gas phases are radially symmetric around the nucleus in some galaxies (J0842 and J0954), others show clear asymmetry and substructure (summarized in Table \ref{tab:AGN_PAH_prop}). A clear example of this are the Pa$\alpha$ maps of J0906 and J1009, where multiple (3+) asymmetric clumps of increased emission are detected more than 1 kpc away from the nucleus. Similar features are seen in the other hydrogen recombination lines, including Pa$\beta$, Br$\alpha$, and Br$\beta$. Although not following the same overall spatial structure, enhanced PAH emission is also present in these regions. Archival \textit{HST}/F606W imaging, mainly tracing the H$\alpha$ emission, of J0906 taken during cycle 22 (PID 13943, P.I.: Reines) provides a broader FoV and shows that these regions are interdispersed throughout the galaxy disk. Put together, these regions are likely pockets of comparatively high star formation activity.

Of the detectable gas phases, the emission structure of the hot H$_2$ molecular gas is the most varied in our sample. In J0906, J0954, and J1009, the distribution of the H$_2$ gas is distinct and does not appear coupled with the other gas phases. This includes clumps and extended emission in regions where there is no notable enhancement in the ionized gas or 3.3$\mu$m PAH emission. Inspection of the rotational and rovibrational line ratios suggest thermal heating as the primary cause of emission in these extended regions. Indeed, previous studies have shown evidence of ionized outflows and radio jets in our sample \citep{Manzano2019,Liu2020,Yang2020}. Specifically, radio observations of J0906 show extended 1.66 GHz emission towards the SW (PA $\sim$ 205$^\circ$), a possible indication of jet activity. While this is in the same direction as the extended H$_2$ gas, the small scales of the radio emission indicate that the jet may only have a limited broader effect.

Similarly, the extended H$_2$ regions in J0954 and J1009 could be caused by the interaction of outflows with the ISM. The most striking case of this is in J1009, where the overall structure of the molecular gas is largely defined as a `plume'-like shape. This emission structure extends 0.5 kpc from the nucleus and is perpendicular to the continuum and PAH emission structure. Moreover, it is oriented along the axis of the blue shifted outflow bicone (see Figure 38 of \cite{Liu2020}). To quantify the significance of this emission, we calculate the gas mass at the edge of the plume, roughly 0.5 kpc from the nucleus. Using a r=0.3$\arcsec$ (see bottom panel of Figure \ref{fig:J1009_maps}) aperture and following the same procedure as described in Section \ref{subsec:results_H2}, we calculate a gas mass of 1.0 M$_\odot$, comparable to that measured in the nucleus. Interestingly, no H$_2$ emission is seen along the semi-major axis of the continuum and PAH emission (see Figures \ref{fig:spectra} and \ref{fig:PAH_maps}). It is thus likely that outflows are playing a key role in heating the H$_2$ gas in this plume.

\subsubsection{Gas Kinematics} \label{subsubsec:gas_kinematics}

The central gas kinematics of dwarfs has remained a focal point of tension between simulations and observations. In the $\Lambda$CDM framework, baryons rotate in a disk with speeds governed by a dark matter halo that is typically modeled after an Navarro-Frenk-White (NFW) distribution \citep{Navarro1996}. While this model accurately describes the gas kinematics within large mass galaxies, inconsistencies arise in the low-mass regime. One of the main issues here is the diversity of rotation curves in the central, kpc-scale regions of dwarf galaxies \citep{Sales2022}. Observed rotation curves show a variety of profiles, with slopes that are steeper, shallower, or even consistent with the NFW profile. So far, proposed baryonic solutions have been unable to reproduce this diversity of rotation profiles and has remained a strong discrepancy between $\Lambda$CDM simulations and observations.

The kinematic maps of our sample (see Figures \ref{fig:J1009_maps}--\ref{fig:J0906_maps}) likewise show diverse gas velocities within our FoV. This is clearly seen in the predominately uniform velocity fields of J1009 and J0842 in contrast to the distinct kinematic gradients of J0954 and J0906, which are most apparent in Pa$\alpha$. While the symmetry in J0906 is consistent with that of a rotating disk, the velocity structure in J0954 appears more varied. We also detect line stratification, where different gas phases show distinct, offset kinematics from each other. This is mainly seen in Pa$\alpha$ and [\ion{Fe}{2}] where differences of up to 100 km s$^{-1}$ are observed, greater than our associated uncertainties of $\sim$20 km s$^{-1}$. Stratification is also seen in J0842 and J0954 but to a lesser degree ($\sim$60 km s$^{-1}$). While our spatial maps depict a diverse set of gas kinematics, it should be noted that differences in the observed extent of each line limits the regions over which we can compare the different gas phases. This particularly true for J0906 and J1009, where [\ion{Fe}{2}] and H$_2$ only cover a fraction of the Pa$\alpha$ extent.

This diversity in gas kinematics is in contrast to the stellar velocity maps presented in \cite{Liu2020}, which generally show velocity gradients that cut across the central regions of the galaxies, consistent with that of a rotating disk. It is interesting to speculate what could cause the gas to have such kinematic diversity, especially when compared to the stellar component. Due to the isolated nature of our sample, it is reasonable to assume that secular processes are a key factor. Due to the strong evidence for AGN in most of our sample, one intriguing possibility is AGN-induced disturbance or feedback. With the presence of outflows in our sample, mechanical feedback could be playing a role in disturbing the local gas. Indeed, previous studies examining the kinematics of dwarfs show increased disturbance in those hosting AGN \citep{Manzano2020}. To further explore this however, larger samples of dwarfs hosting AGN-driven outflows will be needed in order for us ascertain the full effect of AGN on gas kinematics.

\section{Conclusion} \label{sec:Conclusion}

In this article, we present \textit{JWST}/NIRSpec IFU observations of four dwarf galaxies that show strong evidence of AGN activity. Previous observations have shown they all share several common properties, including similar BPT line ratios, broad [\ion{O}{3}] emission, are located in isolated environments, and span a relatively narrow range of stellar masses. Despite these similarities, instead of finding common nuclear conditions, our dataset reveals diverse nuclear properties. We find varied continuum and emission line features, gas excitation, and the spatial morphologies of the multiphase gas across our sample. The key findings of this article are as follows:

\textbullet\; We detect more than 100 unique emission lines across our sample. This includes many fine structure, molecular hydrogen, and hydrogen recombination lines that were not detected previously with ground based facilities. Following inspection of these lines, we find that the flux peaks of the different gas phases are co-spatial and located at the nucleus. Additionally, all four dwarfs in our sample have at least one coronal line detection, with three having more than ten observed, confirming the presence of AGN.

\textbullet\; 3.3$\mu$m PAH emission is detected in the disk regions of all four galaxies. Surprisingly, three of the four in our sample exhibit both PAH and CL emission within the central 100 parsecs of the nucleus. However, PAH suppression is still detected in all four nuclei, as indicated by lower PAH EQWs compared to the outer regions. This suppression is most pronounced in J0906, where no nuclear PAH emission is detected. Determining whether this deficit is due to grain destruction or ionization, however requires observations of the larger PAH grains at MIR wavelengths.

\textbullet\; The continuum shape varies greatly within our sample. In half of our sample, the continuum is relatively flat while the other half rises steeply at wavelengths $>$2$\mu$m. In the latter, fits confirm that a strong contribution from hot dust, likely heated by the AGN, is needed to reproduce the characteristic rising slope. However, we do not find a clear correlation between the continuum slope and the hardness of the radiation field, as estimated through coronal line ratios.

\textbullet\; The temperatures of the hot H$_2$ gas are around 3,000 K for most regions in our sample, and estimates to the masses within the nuclear region range between 1 -- 20 M$_\odot$. A mixture of thermal and non-thermal processes are energizing the molecular hydrogen in the nucleus, as indicated by line ratios and the decoupling of rotational and rovibrational lines in our excitation diagrams. 

\textbullet\; Both symmetric and asymmetric spatial emission features are seen across the different gas phases. The most diverse is H$_2$, where extended emission is detected where no equivalent ionized or PAH emission is observed. The kinematics are also varied in our sample, where both rotationally bound and non-rotation gas is observed, affirming the diversity of rotation curves problem that challenges simulations. Due to the isolated nature of our sample, secular processes such as previously observed AGN-driven outflows and jets could be the cause of these varied kinematics.\\



The expected low luminosity of AGN in dwarf galaxies, coupled with high levels of star formation, make detecting these AGN populations difficult. Moreover, due to the limitations of current survey facilities, the current sample of AGN candidates is biased towards the high mass end ($>$10$^{9.5}$  M$_\odot$) and is therefore not representative of the broader low-mass AGN population. The later half of this decade, however, should prove fruitful for obtaining a more complete census of the low mass AGN population and searching for IMBHs as a whole. New wide field imaging surveys, such as LSST, Roman, and Euclid, are just coming online and will serve as direct upgrades to SDSS. Additionally, the fiber spectroscopy of Subaru/PFS will not only be an improvement in sensitivity and wavelength coverage over SDSS but the smaller fiber size will also allow us to more easily isolate the nuclear emission. These and other contemporary surveys, such as DESI \citep{Darragh2023}, can then be coupled with machine learning algorithms to find trends in large datasets that can identify AGN. One such example are self-organizing maps (SOMs) which can help separate AGN from star forming contaminants \citep{Sanjaripour2025}. The development and synergy of these techniques will be vital in providing a multi-wavelength approach that is necessary to identify the full AGN population.

\begin{acknowledgments}
We thank the referee for their constructive and insightful comments that have helped improve this manuscript. This work is based on observations made with the NASA/ESA/CSA James Webb Space Telescope, and are associated with program $\#$3663. The data were downloaded from the Mikulski Archive for Space Telescopes at the Space Telescope Science Institute, which is operated by the Association of Universities for Research in Astronomy, Inc., under NASA contract NAS 5-03127 for JWST. The \textit{JWST} data used in this paper can be found at doi:\dataset[10.17909/19nr-xp61]{http://dx.doi.org/10.17909/19nr-xp61}. Support for this project was provided by NASA through a grant from the Space Telescope Science Institute, which is operated by the Association of Universities for Research in Astronomy, Inc., under NASA contract NAS 5-03127. This work was financially supported by JSPS KAKENHI grant No. 26K17196 (T.B). T.N. acknowledges funding support from JSPS KAKENHI grant Nos. 23K25911 and 25H00671. MB acknowledges support from the Juan de La Cierva scholarship with reference JDC2023-052684-I, funded by MICIU/AEI/10.13039/501100011033 and FSE+.\\
\end{acknowledgments}

\facilities{JWST(NIRSpec)}

\software{astropy \citep{Astropy2013,Astropy2018,Astropy2022},  JWST Calibration Pipeline \citep{Bushouse2023},
          BADASS \citep{Sexton2021},
         CAFE \citep{DiazSantos2025}}





\appendix

\section{\textsc{CAFE} Fits}

\begin{figure}
\centering
\epsscale{1.0}
\plotone{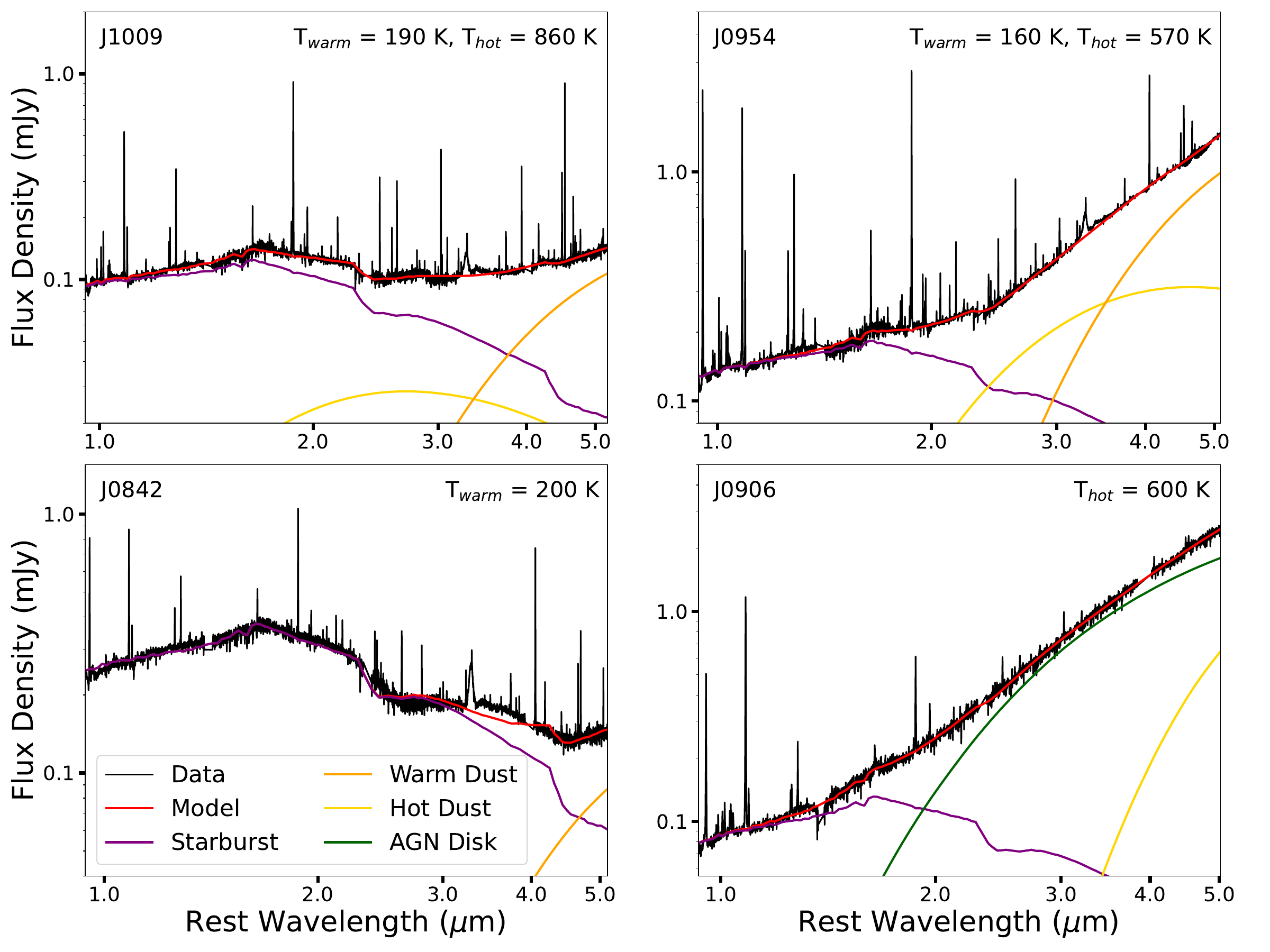}
\caption{A sample of representative \textsc{CAFE} fits to the continuum of each galaxy in our sample. The different components of the best-fit model (starburst, AGN disk, and warm/hot dust) are shown, however most fits do not include all components. The temperatures for the dust component are quoted in the upper right corner of each panel. We emphasize that these values are only representative of the broad contributions of each component and should not be taken at face value. \label{fig:CAFE_fits}}
\end{figure}

The continuum of each galaxy in our sample was fit with CAFE in order to distinguish the relative contributions of the stellar and AGN processes. However, due to the high degree of degeneracy in our fits, these plots are only meant to show the broad trends that emerge after multiple iterations of fitting. In Figure \ref{fig:CAFE_fits}, we plot one such iteration that is representative of these broad trends. For J0906 and J0954, it is clear that starburst alone cannot replicate the steeply rising continuum at longer wavelengths. Some contribution from the AGN, either in the form of hot dust or accretion disk emission, is required. The continuum of J0842 and J1009, on the other hand, show a closer match with the starburst templates. While the contribution from dust in J0842 is likely marginal, warm dust emission makes a more significant contribution at longer wavelengths in J1009.

\section{J0842 [\ion{A\MakeLowercase{r}}{6}] Emission}

\begin{figure}
\centering
\epsscale{1.0}
\plotone{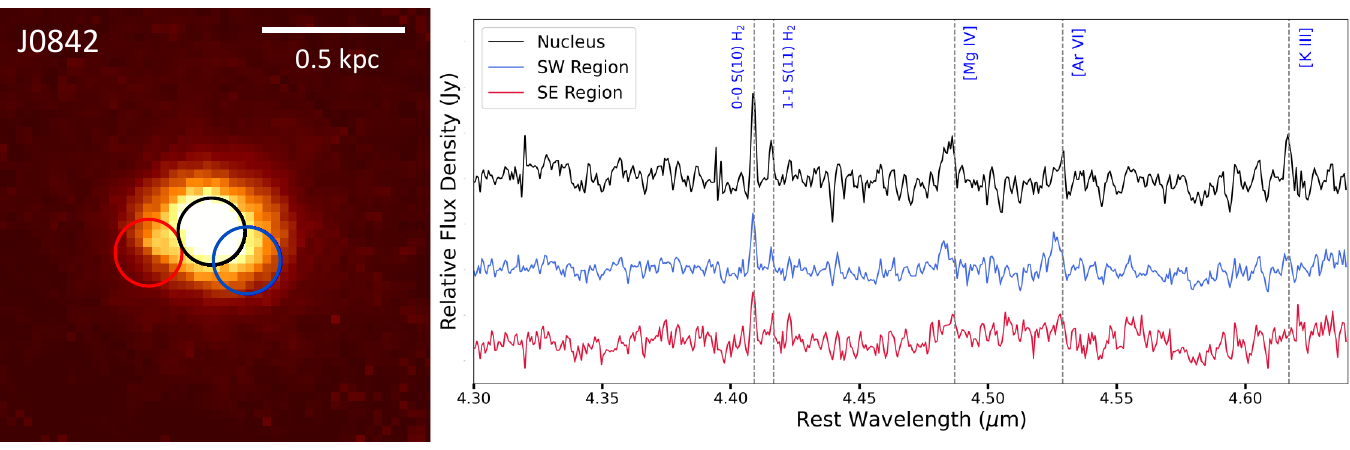}
\caption{(\textit{left}) 3 $\mu$m continuum image of J0842, with circles representing three aperture extractions. (\textit{right}) Spectra of the nucleus, SW and SE regions of J0842, color-coded to match the apertures in left panel. [\ion{Mg}{4}] is detected in both the nucleus and SW region. [\ion{Ar}{6}], on the other hand, is only detected in the SW region.
\label{fig:J0842_ArVI}}
\end{figure}

The analysis of the nuclear spectrum in J0842 only reveals one coronal line detection, [\ion{Mg}{4}] 4.48 $\mu$m (IP = 80 eV). As is discussed in Section \ref{subsubsec:continum_emission_lines}, [\ion{Mg}{4}] emission alone is not a secure tracer of AGN activity since it can be excited by shocks from stellar processes \citep{Pereira2024b}. Closer inspection of regions surrounding the nucleus, however, reveals [\ion{Ar}{6}] (IP = 75 eV) emission detected at $>$3$\sigma$ level $\sim$150 pc southwest of the nucleus (see Figure \ref{fig:J0842_ArVI}). Interestingly, when approaching the nucleus, the [\ion{Ar}{6}] emission decreases and becomes undetectable. While [\ion{Mg}{4}] is also detected in this region, its emission peak is centrally located on the nucleus, suggesting two ionizing sources are involved. No CL emission is observed elsewhere within our FoV. 

The emission line diagnostics of \cite{Costa2026} support an AGN as the excitation source within this SW region. Line ratios of [\ion{Ar}{6}]/[\ion{Mg}{4}] are also consistent with other Seyfert AGN, however the lack of detected Hu-12 emission prevents us from ruling out shock excitation using the models described in \cite{Pereira2024b}. Still, the possibility of an off-nuclear or `wandering AGN' is certainly compelling. To distinguish an off-nuclear AGN from a shock front, however, requires a follow-up analysis that maps out the multi-phase outflows and compares the emission line ratios to shock modeling. 

\section{List of Detected Emission Lines}

The following table lists the emission line fluxes for all four dwarf galaxies. The quoted fluxes are based on a circular (r=$0\farcs3$) aperture centered on the nucleus of each target. Errors were derived from the error extension of the NIRSpec datacubes and the random errors associated with the Bayesian fitting process. For emission lines detected at the 2.5 -- 3.0 $\sigma$ level, we list the measured fluxes as upper limits.

\begin{longtable}{@{\extracolsep{\fill}}cccccc}
\caption{\label{tab:Flux_table} Nuclear Line List and Flux Values}\tabularnewline 
\toprule
\centering
Line & Rest Wavelength & J0842 & J0906 & J0954 & J1009\\
 & ($\mu$m) & \multicolumn{4}{c}{(10$^{-20}$ W m$^{-2}$)}\\
\toprule
\toprule
\endfirsthead

\caption{ Line List and Flux Values (\textit{continued})}\tabularnewline 
\toprule
\centering
Line & Rest Wavelength & J0842 & J0906 & J0954 & J1009\\
 & ($\mu$m) & \multicolumn{4}{c}{(10$^{-20}$ W m$^{-2}$)}\\
\toprule
\toprule
\endhead

\bottomrule
\endfoot

[\ion{S}{3}] Total & 0.9534 & 178.29 $\pm$ 11.15 & 210.86 $\pm$ 3.26 & 538.93 $\pm$ 9.11 & --- \\\relax
(Narrow) & 0.9534 & 92.99 $\pm$ 6.88 & 115.75 $\pm$ 1.96 & 365.85 $\pm$ 6.78 & --- \\\relax
(Broad) & 0.9534 & 85.30 $\pm$ 8.78 & 95.11 $\pm$ 2.61 & 173.08 $\pm$ 6.09 & --- \\\relax
Pa-$\epsilon$ & 0.9549 & 6.74 $\pm$ 3.08 & --- & 19.77 $\pm$ 3.43 & --- \\\relax
[\ion{C}{1}] & 0.9827 & --- & --- & 2.30 $\pm$ 1.32 & --- \\\relax
[\ion{C}{1}] & 0.9853 & --- & 3.47 $\pm$ 1.77 & 7.52 $\pm$ 2.06 & --- \\\relax
[\ion{S}{8}] & 0.9911 & --- & 12.24 $\pm$ 2.87 & 5.99 $\pm$ 2.0 & 5.55 $\pm$ 0.93 \\\relax
Pa-$\delta$ & 1.0052 & 6.26 $\pm$ 1.75 & 8.50 $\pm$ 2.40 & 30.62 $\pm$ 1.80 & 7.65 $\pm$ 0.81 \\\relax
\ion{N}{1} & 1.0112 & --- & 1.83 $\pm$ 1.94 & --- & --- \\\relax
\ion{He}{2} & 1.0126 & --- & 15.4 $\pm$ 2.56 & 15.25 $\pm$ 2.23 & 13.2 $\pm$ 0.88 \\\relax
[\ion{S}{2}] & 1.0290 & 3.86 $\pm$ 2.58 & 8.27 $\pm$ 0.79 & 12.22 $\pm$ 2.22 & 3.18 $\pm$ 1.10 \\\relax
[\ion{S}{2}] & 1.0323 & 6.95 $\pm$ 1.95 & 11.33 $\pm$ 0.84 & 15.88 $\pm$ 2.23 & 3.94 $\pm$ 0.84 \\\relax
[\ion{S}{2}] & 1.0339 & 2.85 $\pm$ 1.71 & 8.20 $\pm$ 0.81 & 10.82 $\pm$ 1.99 & 2.52 $\pm$ 0.74 \\\relax
[\ion{S}{2}] & 1.0373 & --- & 2.05 $\pm$ 0.61 & 4.73 $\pm$ 1.85 & 1.2 $\pm$ 0.16 \\\relax
\ion{N}{1} & 1.0401 & --- & 4.80 $\pm$ 1.39 & 2.18 $\pm$ 1.70 & 0.81 $\pm$ 0.16 \\\relax
\ion{N}{1} & 1.0410 & --- & 4.39 $\pm$ 1.25 & 2.96 $\pm$ 2.04 & --- \\\relax
\ion{He}{1} Total & 1.0833 & 158.86 $\pm$ 8.72 & 538.83 $\pm$ 11.38 & 452.43 $\pm$ 58.48 & 86.84 $\pm$ 5.84 \\\relax
(Narrow) & 1.0833 & 89.04 $\pm$ 6.46 & 305.69 $\pm$ 8.36 & 334.34 $\pm$ 35.54 & 67.09 $\pm$ 4.08 \\\relax
(Broad) & 1.0833 & 69.82 $\pm$ 5.86 & 233.14 $\pm$ 7.73 & 118.09 $\pm$ 46.45 & 19.75 $\pm$ 4.19 \\\relax
Pa-$\gamma$ Total & 1.0941 & 27.27 $\pm$ 4.19 & 18.17 $\pm$ 2.06 & 54.42 $\pm$ 1.49 & 13.63 $\pm$ 0.14 \\\relax
(Narrow) & 1.0941 & 17.86 $\pm$ 3.17 & --- & --- & --- \\\relax
(Broad) & 1.0941 & 9.41 $\pm$ 2.74 & --- & --- & --- \\\relax
[\ion{P}{2}] & 1.1470 & $<$3.71 & --- & 2.05 $\pm$ 0.58 & 1.18 $\pm$ 0.16 \\\relax
\ion{He}{2} & 1.1630 & --- & 2.41 $\pm$ 1.31 & 4.02 $\pm$ 0.73 & 3.23 $\pm$ 0.14 \\\relax
[\ion{P}{2}] & 1.1886 & --- & --- & 5.23 $\pm$ 0.76 & 1.22 $\pm$ 0.14 \\\relax
[\ion{S}{9}] & 1.2530 & --- & 10.90 $\pm$ 1.77 & 5.41 $\pm$ 0.59 & 5.38 $\pm$ 0.12 \\\relax
[\ion{Fe}{2}] Total & 1.2570 & 25.89 $\pm$ 1.62 & 16.02 $\pm$ 1.91 & 62.74 $\pm$ 6.08 & 9.31 $\pm$ 0.12 \\\relax
(Narrow) & 1.2570 & --- & --- & 34.08 $\pm$ 4.12 & --- \\\relax
(Broad) & 1.2570 & --- & --- & 28.66 $\pm$ 4.48 & --- \\\relax
[\ion{Fe}{2}] & 1.2707 & --- & --- & 2.08 $\pm$ 0.64 & $<$0.52 \\\relax
[\ion{Fe}{2}] & 1.2791 & 4.85 $\pm$ 1.41 & 7.87 $\pm$ 1.32 & 7.38 $\pm$ 0.61 & 1.13 $\pm$ 0.14 \\\relax
Pa-$\beta$ Total & 1.2822 & 43.99 $\pm$ 2.24 & 31.39 $\pm$ 1.39 & 116.82 $\pm$ 6.34 & 32.05 $\pm$ 5.75 \\\relax
(Narrow) & 1.2822 & 32.54 $\pm$ 1.44 & --- & 78.47 $\pm$ 4.71 & 25.75 $\pm$ 4.06 \\\relax
(Broad) & 1.2822 & 11.45 $\pm$ 1.72 & --- & 38.35 $\pm$ 4.24 & 6.30 $\pm$ 4.07 \\\relax
\ion{O}{1} & 1.3167 & $<$2.58 & --- & 2.47 $\pm$ 0.52 & --- \\\relax
[\ion{Fe}{2}] & 1.3209 & 6.58 $\pm$ 1.61 & 4.34 $\pm$ 1.60 & 15.05 $\pm$ 0.67 & 2.60 $\pm$ 0.11 \\\relax
[\ion{Fe}{2}] & 1.3724 & 4.90 $\pm$ 1.29 & --- & 9.47 $\pm$ 0.64 & 1.91 $\pm$ 0.11 \\\relax
[\ion{Si}{10}] & 1.4305 & --- & 7.12 $\pm$ 0.98 & 3.12 $\pm$ 0.58 & --- \\\relax
\ion{N}{1} & 1.4761 & --- & --- & $<$1.87 & $<$0.92 \\\relax
[\ion{Fe}{2}] & 1.5339 & --- & --- & 5.52 $\pm$ 0.47 & --- \\\relax
Br-13 & 1.6114 & --- & --- & $<$3.48 & --- \\\relax
Br-12 & 1.6412 & --- & --- & $<$2.46 & --- \\\relax
[\ion{Fe}{2}] Total & 1.6435 & 21.40 $\pm$ 6.13 & 19.07 $\pm$ 1.08 & 51.57 $\pm$ 2.72 & 7.50 $\pm$ 0.07 \\\relax
(Narrow) & 1.6435 & 9.57 $\pm$ 3.63 & --- & 27.46 $\pm$ 1.96 & --- \\\relax
(Broad) & 1.6435 & 11.82 $\pm$ 4.94 & --- & 24.12 $\pm$ 1.89 & --- \\\relax
[\ion{Fe}{2}] & 1.6773 & --- & --- & 6.60 $\pm$ 0.43 & --- \\\relax
Br-11 & 1.6811 & --- & --- & 4.70 $\pm$ 0.37 & --- \\\relax
1--0 S(8) H$_2$ & 1.7147 & --- & --- & 2.84 $\pm$ 0.4 & --- \\\relax
Br-10 & 1.7367 & --- & --- & 6.27 $\pm$ 0.52 & 1.32 $\pm$ 0.06 \\\relax
[\ion{Fe}{2}] & 1.8094 & 6.22 $\pm$ 1.15 & --- & 9.65 $\pm$ 0.46 & 2.51 $\pm$ 0.11 \\\relax
Br-$\epsilon$ & 1.8179 & 3.67 $\pm$ 0.98 & --- & 8.45 $\pm$ 0.45 & 3.20 $\pm$ 0.12 \\\relax
1--0 S(5) H$_2$ & 1.8358 & 5.79 $\pm$ 0.96 & --- & 3.01 $\pm$ 0.43 & --- \\\relax
\ion{He}{2} & 1.8641 & --- & 6.79 $\pm$ 1.16 & 5.42 $\pm$ 0.58 & 5.19 $\pm$ 0.34 \\\relax
\ion{He}{1} & 1.8691 & 3.65 $\pm$ 0.96 & --- & 9.08 $\pm$ 0.56 & 2.73 $\pm$ 0.34 \\\relax
\ion{He}{1} & 1.8702 & --- & --- & 2.59 $\pm$ 0.54 & 1.17 $\pm$ 0.33 \\\relax
Pa-$\alpha$ Total & 1.8756 & 94.64 $\pm$ 19.04 & 101.11 $\pm$ 7.32 & 311.89 $\pm$ 9.72 & 84.00 $\pm$ 3.61 \\\relax
(Narrow) & 1.8756 & 51.02 $\pm$ 14.32 & 52.32 $\pm$ 5.39 & 183.83 $\pm$ 7.67 & 61.55 $\pm$ 2.61 \\\relax
(Broad 1) & 1.8756 & 43.62 $\pm$ 12.54 & 48.79 $\pm$ 4.96 & 94.39 $\pm$ 4.50 & 22.45 $\pm$ 2.50 \\\relax
(Broad 2) & 1.8756 & --- & --- & 33.67 $\pm$ 3.93 & --- \\\relax
[\ion{S}{11}] & 1.9220 & --- & 6.82 $\pm$ 0.38 & --- & 1.77 $\pm$ 0.10 \\\relax
Br-$\delta$ & 1.9451 & 3.32 $\pm$ 0.76 & --- & 12.51 $\pm$ 0.38 & 2.40 $\pm$ 0.06 \\\relax
1--0 S(3) H$_2$ & 1.9576 & 9.86 $\pm$ 0.80 & --- & 6.02 $\pm$ 0.40 & 1.59 $\pm$ 0.10 \\\relax
[\ion{Si}{6}] Total & 1.9630 & --- & 28.71 $\pm$ 1.12 & 15.48 $\pm$ 1.61 & 8.18 $\pm$ 0.08 \\\relax
(Narrow) & 1.9630 & --- & --- & 11.01 $\pm$ 1.40 & --- \\\relax
(Broad) & 1.9630 & --- & --- & 4.47 $\pm$ 0.80 & --- \\\relax
1--0 S(2) H$_2$ & 2.0337 & 3.39 $\pm$ 0.79 & --- & 2.34 $\pm$ 0.42 & $<$0.50 \\\relax
\ion{He}{1} & 2.0581 & 4.16 $\pm$ 0.73 & --- & 11.38 $\pm$ 0.34 & 1.05 $\pm$ 0.06 \\\relax
2--1 S(3) H$_2$ & 2.0735 & --- & --- & 0.80 $\pm$ 0.31 & --- \\\relax
1--0 S(1) H$_2$ & 2.1218 & 8.47 $\pm$ 0.68 & 5.36 $\pm$ 0.77 & 6.19 $\pm$ 0.31 & 1.47 $\pm$ 0.06 \\\relax
Br-$\gamma$ Total & 2.1661 & 7.73 $\pm$ 0.71 & 8.06 $\pm$ 0.89 & 24.24 $\pm$ 3.61 & 5.77 $\pm$ 0.07 \\\relax
(Narrow) & 2.1661 & --- & --- & 15.62 $\pm$ 2.40 & --- \\\relax
(Broad) & 2.1661 & --- & --- & 8.62 $\pm$ 2.70 & --- \\\relax
\ion{He}{2} & 2.1891 & --- & --- & --- & 0.63 $\pm$ 0.06 \\\relax
1--0 S(0) H$_2$ & 2.2235 & 2.67 $\pm$ 0.56 & --- & 2.78 $\pm$ 0.36 & 0.66 $\pm$ 0.05 \\\relax
2--1 S(1) H$_2$ & 2.2477 & 1.57 $\pm$ 0.64 & $<$2.94 & 1.66 $\pm$ 0.33 & 0.69 $\pm$ 0.06 \\\relax
[\ion{Ca}{8}] & 2.3218 & --- & --- & 2.15 $\pm$ 0.6 & 1.58 $\pm$ 0.06 \\\relax
1--0 Q(1) H$_2$ & 2.4065 & 7.09 $\pm$ 0.47 & --- & 6.16 $\pm$ 0.24 & --- \\\relax
1--0 Q(2) H$_2$ & 2.4134 & $<$1.23 & --- & 1.67 $\pm$ 0.23 & --- \\\relax
1--0 Q(3) H$_2$ & 2.4237 & 5.74 $\pm$ 0.46 & 3.81 $\pm$ 0.53 & 4.65 $\pm$ 0.23 & 1.00 $\pm$ 0.05 \\\relax
1--0 Q(5) H$_2$ & 2.4547 & 3.07 $\pm$ 0.47 & --- & 2.27 $\pm$ 0.31 & $<$0.54 \\\relax
[\ion{Si}{7}] Total & 2.4833 & --- & 24.36 $\pm$ 0.61 & 22.16 $\pm$ 1.82 & 12.85 $\pm$ 0.05 \\\relax
(Narrow) & 2.4833 & --- & --- & 12.63 $\pm$ 1.22 & --- \\\relax
(Broad) & 2.4833 & --- & --- & 9.53 $\pm$ 1.36 & --- \\\relax
Pf-17 & 2.4953 & --- & --- & $<$1.23 & --- \\\relax
1--0 Q(7) H$_2$ & 2.4999 & --- & 4.44 $\pm$ 0.22 & $<$0.85 & --- \\\relax
Pf-16 & 2.5261 & --- & --- & $<$1.07 & --- \\\relax
2--1 Q(1) H$_2$ & 2.5510 & $<$1.51 & --- & $<$1.13 & $<$0.60 \\\relax
2--1 Q(2) H$_2$ & 2.5585 & --- & --- & $<$1.09 & $<$0.69 \\\relax
Pf-15 & 2.5643 & --- & --- & $<$0.96 & --- \\\relax
[\ion{Si}{9}] & 2.5839 & --- & --- & 4.43 $\pm$ 0.52 & 4.74 $\pm$ 0.05 \\\relax
Pf-14 & 2.6127 & --- & --- & 1.99 $\pm$ 0.35 & $<$0.40 \\\relax
Br-$\beta$ Total & 2.6259 & 14.66 $\pm$ 0.59 & 15.44 $\pm$ 0.58 & 49.32 $\pm$ 2.26 & 13.36 $\pm$ 1.61 \\\relax
(Narrow) & 2.6259 & --- & --- & 25.34 $\pm$ 1.79 & 7.96 $\pm$ 0.94 \\\relax
(Broad) & 2.6259 & --- & --- & 23.99 $\pm$ 1.38 & 5.40 $\pm$ 1.30 \\\relax
2--1 Q(7) H$_2$ & 2.6539 & --- & --- & $<$0.57 & --- \\\relax
Pf-13 & 2.6751 & --- & --- & 1.53 $\pm$ 0.22 & --- \\\relax
3--2 Q(1) H$_2$ & 2.7103 & --- & --- & $<$0.75 & --- \\\relax
Pf-12 & 2.7583 & --- & --- & 2.02 $\pm$ 0.20 & 0.74 $\pm$ 0.04 \\\relax
2--1 O(2) H$_2$ & 2.7862 & --- & --- & --- & $<$0.42 \\\relax
1--0 O(3) H$_2$ & 2.8025 & 5.77 $\pm$ 0.38 & --- & 5.16 $\pm$ 0.17 & 1.22 $\pm$ 0.03 \\\relax
3--2 Q(7) H$_2$ & 2.8250 & --- & --- & --- & 0.92 $\pm$ 0.05 \\\relax
Pf-11 & 2.8730 & --- & --- & 2.96 $\pm$ 0.21 & 0.74 $\pm$ 0.03 \\\relax
\ion{O}{1} & 2.8933 & --- & --- & $<$0.87 & --- \\\relax
2--1 O(3) H$_2$ & 2.9741 & 1.03 $\pm$ 0.32 & --- & 1.15 $\pm$ 0.24 & 0.66 $\pm$ 0.03 \\\relax
1--0 O(4) H$_2$ & 3.0039 & 2.29 $\pm$ 0.56 & --- & 1.90 $\pm$ 0.23 & 0.54 $\pm$ 0.04 \\\relax
[\ion{Mg}{8}] Total & 3.0276 & --- & 33.78 $\pm$ 0.59 & 23.68 $\pm$ 3.26 & 15.46 $\pm$ 2.29 \\\relax
(Narrow) & 3.0276 & --- & --- & 10.34 $\pm$ 2.21 & 11.79 $\pm$ 1.67 \\\relax
(Broad) & 3.0276 & --- & --- & 13.33 $\pm$ 2.40 & 3.67 $\pm$ 1.58 \\\relax
Pf-$\epsilon$ & 3.0392 & 1.11 $\pm$ 0.55 & --- & 4.68 $\pm$ 0.41 & 1.26 $\pm$ 0.04 \\\relax
\ion{He}{2} & 3.0917 & --- & --- & 3.44 $\pm$ 0.73 & 2.34 $\pm$ 0.04 \\\relax
3--2 O(3) H$_2$ & 3.1638 & --- & --- & --- & $<$0.24 \\\relax
2--1 O(4) H$_2$ & 3.1898 & --- & --- & 0.80 $\pm$ 0.25 & 0.41 $\pm$ 0.06 \\\relax
[\ion{Ca}{4}] & 3.2071 & --- & 14.82 $\pm$ 0.54 & 9.38 $\pm$ 0.28 & 2.10 $\pm$ 0.04 \\\relax
1--0 O(5) H$_2$ & 3.2353 & 3.17 $\pm$ 0.45 & --- & 3.23 $\pm$ 0.47 & 0.59 $\pm$ 0.34 \\\relax
3.3$\mu$m PAH & 3.2890 & 104.53 $\pm$ 4.01 & --- & 218.88 $\pm$ 5.72 & 52.11 $\pm$ 6.14 \\\relax
Pf-$\delta$ & 3.2970 & 2.53 $\pm$ 0.59 & --- & 7.91 $\pm$ 0.56 & 1.92 $\pm$ 0.54 \\\relax
2--1 O(5) H$_2$ & 3.4379 & $<$0.47 & --- & --- & $<$0.16 \\\relax
1--0 O(6) H$_2$ & 3.5008 & 0.85 $\pm$ 0.38 & --- & 0.97 $\pm$ 0.22 & --- \\\relax
\ion{He}{1} & 3.5445 & --- & --- & --- & $<$0.16 \\\relax
0--0 S(15) H$_2$ & 3.6262 & 0.91 $\pm$ 0.39 & --- & 0.77 $\pm$ 0.22 & $<$0.19 \\\relax
[\ion{Al}{6}] & 3.6598 & --- & --- & 4.59 $\pm$ 0.27 & 1.59 $\pm$ 0.04 \\\relax
2--1 O(6) H$_2$ & 3.7237 & $<$0.35 & --- & --- & --- \\\relax
Pf-$\gamma$ & 3.7410 & 3.63 $\pm$ 0.44 & --- & 10.14 $\pm$ 0.20 & 2.60 $\pm$ 0.04 \\\relax
Humph-17 & 3.7494 & --- & --- & --- & $<$0.17 \\\relax
1--0 O(7) H$_2$ & 3.8074 & 1.01 $\pm$ 0.35 & --- & 0.88 $\pm$ 0.15 & 0.24 $\pm$ 0.05 \\\relax
Humph-16 & 3.8195 & --- & --- & --- & 0.33 $\pm$ 0.04 \\\relax
0--0 S(13) H$_2$ & 3.8461 & 1.48 $\pm$ 0.39 & --- & 0.80 $\pm$ 0.19 & $<$0.18 \\\relax
Humph-15 & 3.9075 & --- & --- & --- & 0.33 $\pm$ 0.04 \\\relax
[\ion{Si}{9}] Total & 3.9357 & --- & --- & --- & 9.88 $\pm$ 1.34 \\\relax
(Narrow) & 3.9357 & --- & --- & --- & 8.31 $\pm$ 0.89 \\\relax
(Broad) & 3.9357 & --- & --- & --- & 1.57 $\pm$ 0.99 \\\relax
Br-$\alpha$ Total & 4.0510 & 30.35 $\pm$ 1.17 & 28.31 $\pm$ 0.55 & 88.35 $\pm$ 1.33 & --- \\\relax
(Narrow) & 4.0510 & 15.16 $\pm$ 0.69 & --- & 50.93 $\pm$ 0.86 & --- \\\relax
(Broad) & 4.0510 & 15.20 $\pm$ 0.94 & --- & 37.41 $\pm$ 1.02 & --- \\\relax
[\ion{Ca}{7}] & 4.0864 & --- & --- & 1.59 $\pm$ 0.27 & 0.86 $\pm$ 0.08 \\\relax
[\ion{Ca}{5}] & 4.1594 & --- & 6.68 $\pm$ 1.32 & 7.14 $\pm$ 0.30 & 2.47 $\pm$ 0.06 \\\relax
Humph-13 & 4.1708 & --- & --- & 1.08 $\pm$ 0.19 & 0.29 $\pm$ 0.06 \\\relax
0--0 S(11) H$_2$ & 4.1811 & 3.10 $\pm$ 0.30 & --- & 1.47 $\pm$ 0.20 & 0.28 $\pm$ 0.06 \\\relax
\ion{He}{1} & 4.2954 & --- & --- & 1.97 $\pm$ 0.20 & 0.48 $\pm$ 0.05 \\\relax
Humph-12 & 4.3765 & --- & --- & 1.33 $\pm$ 0.30 & 0.49 $\pm$ 0.05 \\\relax
0--0 S(10) H$_2$ & 4.4091 & 1.09 $\pm$ 0.24 & --- & 0.84 $\pm$ 0.19 & --- \\\relax
1--1 S(11) H$_2$ & 4.4166 & $<$0.45 & --- & --- & --- \\\relax
[\ion{Mg}{4}] Total & 4.4871 & 1.08 $\pm$ 0.37 & 30.18 $\pm$ 1.43 & 20.16 $\pm$ 1.87 & 6.32 $\pm$ 0.08 \\\relax
(Narrow) & 4.4871 & --- & --- & 17.09 $\pm$ 1.64 & --- \\\relax
(Broad) & 4.4871 & --- & --- & 3.07 $\pm$ 0.90 & --- \\\relax
[\ion{Ar}{6}] Total & 4.5291 & --- & 29.78 $\pm$ 1.77 & 27.21 $\pm$ 0.55 & 24.37 $\pm$ 2.55 \\\relax
(Narrow) & 4.5291 & --- & --- & --- & 19.98 $\pm$ 1.78 \\\relax
(Broad) & 4.5291 & --- & --- & --- & 4.39 $\pm$ 1.82 \\\relax
[\ion{K}{3}] & 4.6168 & 0.71 $\pm$ 0.27 & --- & --- & 0.66 $\pm$ 0.32 \\\relax
Pf-$\beta$ Total & 4.6540 & 5.04 $\pm$ 0.33 & --- & 16.84 $\pm$ 0.58 & 4.30 $\pm$ 0.87 \\\relax
(Narrow) & 4.6540 & --- & --- & --- & 3.32 $\pm$ 0.59 \\\relax
(Broad) & 4.6540 & --- & --- & --- & 0.98 $\pm$ 0.65 \\\relax
Humph-$\epsilon$ & 4.6725 & 0.76 $\pm$ 0.29 & --- & --- & 0.68 $\pm$ 0.08 \\\relax
[\ion{Na}{7}] & 4.6834 & --- & --- & --- & 1.12 $\pm$ 0.08 \\\relax
0--0 S(9) H$_2$ & 4.6946 & 6.62 $\pm$ 0.25 & --- & 3.75 $\pm$ 0.52 & 0.55 $\pm$ 0.06 \\\relax
[\ion{Si}{10}] & 4.7633 & --- & --- & --- & 1.06 $\pm$ 0.06 \\\relax
0--0 S(8) H$_2$ & 5.0531 & 3.20 $\pm$ 0.22 & --- & ---  & $<$0.35 \\\relax
Humph-$\delta$ & 5.1287 & --- & --- & --- & 0.78 $\pm$ 0.06 \\
\hline
\multicolumn{6}{p{16.0cm}}{\textsc{Note}---Columns: (1) Emission line. If multiple components were fit, separate fluxes for the narrow, broad and total are listed. (2) Rest-frame wavelength. (3) - (6) Fluxes, as derived from a r=$0\farcs3$ nuclear extraction. They are aperture corrected but not extinction corrected. Units are in 10$^{-20}$ W m$^{-2}$. Upper limits are provided if the S/N of the line is less than 3 but greater than 2.5.}
\end{longtable}

\pagebreak


\end{document}